\magnification \magstep1
\input amssym.def
\input amssym.tex
\centerline{\bf Quantum Amplitudes in Black--Hole Evaporation:}
\centerline{\bf Complex Approach and Spin--0 Amplitude}
\bigskip
\centerline{A.N.St.J.Farley and P.D.D'Eath}
\bigskip
\smallskip
\centerline{Department of Applied Mathematics and Theoretical Physics,} 
\smallskip
\centerline{Centre for Mathematical Sciences,{~}University of Cambridge,}
\smallskip
\centerline{Wilberforce Road,{~}Cambridge CB3 0WA,{~}United Kingdom} 
\bigskip
\centerline{Abstract}
\medskip
This paper is concerned with the quantum--mechanical decay 
of a Schwarzschild--like black hole, 
formed by gravitational collapse, 
into almost--flat space--time and weak radiation at a very late time.  
We evaluate quantum amplitudes 
(not just probabilities) 
for transitions from initial to final states.  
This quantum description shows that no information is lost in collapse 
to a black hole.  
Boundary data for the gravitational field and 
(in this paper) 
a scalar field are posed on an initial space--like hypersurface 
${\Sigma}_{I}$ 
and a final surface 
${\Sigma}_{F}{\;}$.  
These asymptotically--flat 3--surfaces 
are separated by a Lorentzian proper--time interval 
$T$ 
(typically very large), 
as measured at spatial infinity.  
The boundary--value problem is made well--posed,
both classically and quantum--mechanically, 
by a rotation of 
$T$ 
into the lower--half complex plane: 
$T\longrightarrow{\mid}T{\mid}\exp(-{\,}i{\theta})$, 
with 
$0<{\theta}\leq{{\pi}/2}{\;}$. 
This corresponds to Feynman's 
$+{\,}i{\epsilon}$ 
prescription.  
We consider the classical boundary--value problem 
and calculate the second--variation classical Lorentzian action
$S^{(2)}_{\rm class}$
as a functional of the boundary data.
Following Feynman,
the Lorentzian quantum amplitude is recovered in the limit 
${\theta}\longrightarrow{0}_{+}$ 
from the well--defined complex--$T$ amplitude.
Dirac's canonical approach to the quantisation of constrained systems 
shows that,
for locally--supersymmetric theories of gravity,
the amplitude is exactly semi--classical,
namely,
$\exp(iS^{(2)}_{\rm class})$
for weak perturbations,
apart from delta--functionals of the supersymmetry constraints.
We treat such quantum amplitudes for weak scalar--field configurations on 
${\Sigma}_{F}{\,}$, 
taking 
(for simplicity) 
the weak final gravitational field to be spherically symmetric.  
The treatment involves adiabatic solutions of the scalar wave equation.
This considerably extends work reported in previous papers,
by giving explicit expressions for the real and imaginary parts 
of such quantum amplitudes.\par
\par
\medskip
\noindent
{\bf 1. Introduction}
\medskip
\indent
This paper is concerned with the simplest example of quantum radiation 
following gravitational collapse to a black hole, 
in which the Lagrangian contains only Einstein gravity 
and a minimally--coupled massless scalar field 
${\phi}{\,}$.  
As with emitted particles of spin 
$s>0$ 
[1--3], 
so here we find that the emission process with 
$s=0$ 
may be described in terms of quantum amplitudes
(not just probabilities) 
[1,4--6].  
An effective--field approach to this question 
has also been described{~}---{~}see [7,8].\par
\smallskip
\indent
We compute such amplitudes;  
the basic quantity is the amplitude to go from data 
on an initial asymptotically--flat space--like hypersurface 
${\Sigma}_{I}$ 
to data on a final surface 
${\Sigma}_{F}{\;}$.  
The proper--time separation between 
${\Sigma}_{I}$ 
and 
${\Sigma}_{F}{\;}$, 
as measured at spatial infinity, 
is 
$T{\,}$.
The 4--dimensional space--time metric is denoted by 
$g_{{\mu}{\nu}}{\;}{\,}({\mu},{\nu}=0,1,2,3)$.  
The gravitational data to be specified on 
${\Sigma}_{I}$ 
and on 
${\Sigma}_{F}$ 
(regarded as surfaces of constant coordinate 
$x^{0}$)
are the intrinsic positive--definite spatial 3--metric 
$h_{ij}
=
g_{ij}$ 
on each 3--surface, 
written as 
$h_{ijI}$ 
and 
$h_{ijF}{\,}$.  
The scalar data are taken to be the scalar field 
${\phi}_{I}$ 
on 
${\Sigma}_{I}$
and 
${\phi}_{F}$ 
on 
${\Sigma}_{F}{\;}$.
\par
\smallskip
\indent 
In field theory, 
quantum calculations are usually carried out about some classical solution.  
The above boundary data with 
$T$ 
real, 
however, 
are clearly ill--suited to such a classical boundary--value calculation, 
since classical hyperbolic 
(wave--like) 
equations are, 
in general, 
badly posed when subject to data on an outer boundary [9].
(The natural arena for hyperbolic equations 
is in the Cauchy evolution problem [10,11].)  
A simple example of such badly--posed behaviour is given in [5] 
(see also Sec.2.4 of [12]), 
which treats the 1--space, 
1--time weak--field analogue of the boundary--value problem 
of the previous paragraph.  
In this example, 
there is only a real massless scalar field in flat Minkowski space--time.  
The r\^oles of 
${\Sigma}_{I}$ 
and 
${\Sigma}_{F}$ 
are taken by the pair of lines 
$\{t=0{\,},T{\;};{\,}-{\,}{\infty}<x<{\infty}\}$, 
and, 
for simplicity, 
data for 
${\phi}$ 
are chosen with 
${\phi}(t=0{\,},x)=0{\,},
{\,}{\phi}(t=T,x)={\phi}_{1}(x)$, 
where 
${\phi}_{1}$ 
is a specified function of rapid decrease as 
${\mid}x{\mid}\longrightarrow{\infty}{\,}$.   
One can see from [5] that this boundary--value problem becomes well--posed, 
once 
$T$ 
is rotated into the lower--half complex plane:
$$T
{\;}
\longrightarrow
{\;}
{\mid}T{\mid}\exp(-{\,}i{\theta})
{\qquad}{\qquad}{\qquad}{\quad}
(0{\,}<{\,}{\theta}{\,}\leq{\,}{\pi}/2).
\eqno(1.1)$$
\noindent 
However, 
the limit 
${\theta}\longrightarrow{0}$ 
is somewhat singular.  
This fits in with the behaviour that one might expect:  
for 
${\,}{\theta}{\,}$ 
exactly equal to 
${\,}{\pi}/2{\;}$, 
one has a real elliptic boundary--value problem for the Laplace equation, 
which is known to be well--posed.  
In between, 
for 
$0<{\theta}<{\pi}/2{\;}$, 
the differential equation has complex coefficients 
but the boundary--value problem is still well--posed, 
with existence, 
uniqueness and analyticity of the solutions.  
We are, 
in fact, 
viewing a 
{\it strongly elliptic} 
boundary--value problem when
$0<{\theta}\leq{\pi}/2$ 
[13].
For
${\theta}
=
0{\,}$,
one is,
as above,
trying to fit a wave--like system into a boundary--value problem.\par
\smallskip
\indent
Correspondingly, 
in our field--theoretic black--hole evaporation problem, 
we make the same complex rotation (1.1) of 
$T{\,}$.  
Weak--field analyses suggest that, 
with such a complex rotation, 
the classical Dirichlet boundary--value problem 
for a perturbative scalar field 
${\phi}$ 
(and also for a linearised gravitational field 
${\delta}g_{{\mu}{\nu}}$) 
become well--posed.  
Hence, 
for an evolution of the black hole 
which does not deviate too far from the spherical, 
one expects to be able to study a semi--classical expansion of the amplitude,
provided
${\rm Im}(T){\;}{\leq}{\;}0{\,}$.  
Once one has found the quantum amplitude for a given value of  
${\theta}{\,}$, 
with 
$0<{\theta}{\;}{\leq}{\;}{\pi}/2{\;}$, 
one can rotate 
${\,}{\theta}{\,}$
back towards zero,
following Feynman's 
${\,}+{\,}i{\epsilon}{\,}$ 
prescription [14,15].   
In the context of scalar--field evaporation from black holes,  
this procedure will be treated in Sec.8 below.
The slight complexification of 
${\,}T{\,}$ 
induces an imaginary part in the total Lorentzian action, 
which is crucial in computing the appropriate quantum amplitude.  
Conversely, 
even for fairly small values of
${\,}{\theta}{\,}$, 
solution of the rotated classical boundary--value problem 
is expected to smooth any variations or oscillations of the boundary data, 
as one moves from the boundary
${\,}{\Sigma}_{I}{\,}$
or 
${\,}{\Sigma}_{F}{\,}$
into the interior by a few multiples of the relevant wavelength.\par
\smallskip
\indent
To fix one's physical intuition, 
imagine that the initial distribution of scalar field 
${\phi}_{I}$ 
on 
${\Sigma}_{I}$ 
is nearly spherically--symmetric and extremely diffuse, 
with almost all of the mass distributed over radii much greater than the 
'Schwarzschild radius' 
$2M_{0}{\;}$.  
Here, 
$M_{0}$ 
is the 
$ADM$ 
(Arnowitt--Deser--Misner) 
mass, 
which is defined in terms of the rate of fall--off of the initial data, 
at large radii on a given spacelike hypersurface [16].  
The initial 3--metric 
$h_{ij}$ 
on 
${\Sigma}_{I}$ 
will be almost spherically symmetric, 
and will vary very slowly with radius.
The final surface 
${\Sigma}_{F}$ 
will be chosen at very late 
$T{\,}$, 
so as to register all the evaporated radiation.  
The total 
$ADM$ 
mass on the final surface must also equal 
$M_{0}{\;}$, 
since otherwise the classical boundary--value problem with a given 
(finite) 
time--interval
{~}$T$ 
at spatial infinity will have no solution [17,18].  
As regards the classical solution in the interior 
$(0<t<T)$, 
the geometry is well approximated at late times 
by the radiating Schwarzschild--like Vaidya metric [19--21].  
The classical scalar--field solution will depend on the enormous amount 
of detail which, 
in general, 
is present in the prescribed final data 
${\phi}_{F}$ 
on 
${\Sigma}_{F}{\;}$.
\par
\smallskip
\indent
In Sec.2, 
we comment on the need to work, 
not just with the Einstein/scalar Lagrangian, 
but with supergravity or with gauge--invariant supergravity models 
which include supermatter, 
in order that quantum amplitudes should be meaningful.  
Indeed,
in the locally--supersymmetric case,
for a large class of models,
the amplitude above turns out to be exactly semi--classical,
in a certain sense{~}---{~}see Eq.(2.2) below [12,22--24].  
In Sec.3, 
we discuss the 
'background' 
nearly--spherically--symmetric 4--metric 
${\gamma}_{{\mu}{\nu}}{\;}$, 
needed in a self--consistent treatment of the classical field equations.  
In particular, 
the Einstein field equations give, 
at lowest order, 
a 
'source' 
for 
${\gamma}_{{\mu}{\nu}}$
which includes the energy--momentum tensor of the scalar field 
${\phi}{\,}$,
together with a source quadratic in graviton perturbations 
(as well as corresponding sources 
for any other matter fields present).  
Sec.4 treats the decomposition of scalar perturbations in spherical harmonics, 
assuming that the background 4--metric
${\gamma}_{{\mu}{\nu}}$ 
is spherically symmetric.  
In Sec.5, 
we describe the classical action functionals 
$S_{\rm class}$ 
(Lorentzian action) 
or 
$I_{\rm class}$ 
(Euclidean or Riemannian action), 
related by 
$iS_{\rm class}
=
-{\,}I_{\rm class}{\,}$, 
for the Einstein/scalar system.  
Because 
$S_{\rm class}$ 
or 
$I_{\rm class}$ 
are evaluated at a solution of the classical field equations, 
they reduce to a sum of boundary terms.  
Sec.6 treats the adiabatic radial functions for evolution of the 
(linearised) 
scalar field, 
based on the general treatment in Sec.4.   
For adiabatic perturbations
(high frequencies), 
the time--dependence is approximately harmonic and can be factored out, 
leading to a second--order radial equation for given frequency
${\omega}$
and angular quantum numbers
${\ell}{\,},
m{\,}$.
In Sec.6,
we also describe the 
'coordinates' 
which are most convenient in specifying the final data 
${\phi}_{F}$ 
on 
${\Sigma}_{F}{\;}$.  
A suitable basis of radial eigenfunctions on the final surface 
${\Sigma}_{F}$ 
is discussed in Sec.7.  
The analytic continuation process, 
in which a Lorentzian quantum amplitude is derived from the limit 
${\theta}\longrightarrow{0}_{+}$ 
of the amplitude for complex time--separation 
$T{\,}$,
as in Eq.(1.1) [14,15], 
is treated in Sec.8.  
By these methods, 
one can, 
if desired, 
evaluate both the real and imaginary parts 
of the lowest--order perturbative classical action 
$S^{(2)}_{\rm class}{\;}$, 
and hence of the semi--classical amplitude 
$\exp(iS^{(2)}_{\rm class})$.
Sec.9 contains the Conclusion.\par
\medskip
\noindent
{\bf 2. The quantum amplitude for bosonic boundary data}
\medskip
\indent
Consider, 
at present, 
the 
'Euclidean' 
quantum amplitude to go between prescribed initial and final purely bosonic 
(gravitational and matter) 
data, 
on a pair of 3--surfaces, 
each 
'topologically' 
(diffeomorphically) 
${\Bbb R}^{3}$ 
and each carrying an asymptotically--flat 3--metric.  
One further needs to specify the proper 
(Euclidean) 
distance 
${\tau}{\,}$, 
measured orthogonally between the two surfaces at spatial infinity.  
From one
(purely formal)
point of view,
this amplitude can be regarded as given by a Feynman path integral 
over all Riemannian infilling 4--geometries, 
together with any other fields, 
each such configuration being weighted by 
$\exp(-{\,}I)$, 
where 
$I$ 
is the 
'Euclidean action' 
of the configuration.   
If this definition were meaningful, 
one would expect that the resulting 
'Euclidean' 
quantum amplitude would have the semi--classical form
$${\rm Amplitude}
{\;}{\;}{\,}
\sim
{\;}{\;}{\,}
\bigl(A_{0}{\,}+{\,}{\hbar}A_{1}{\,}
+{\,}{\hbar}^{2}A_{2}
+{\,}{\ldots}{\;}\bigr)
\exp\bigl(-{\,}I_{B}/{\hbar}\bigr),
\eqno(2.1)$$
\noindent
giving an asymptotic expansion in the limit that 
$(I_{B}/{\hbar}){\,}{\longrightarrow}{\,}0{\,}$.
Here, 
$I_{B}$ 
is the classical 
'Euclidean action' 
of a Riemannian solution of the coupled Einstein 
and matter classical field equations, 
subject to the boundary conditions.  
For simplicity, 
we assume that there is a unique classical solution, 
up to gauge and coordinate transformations.  
It is quite feasible, 
though, 
in certain theories and for certain boundary data, 
to have instead 
(say) 
a complex--conjugate pair of classical solutions [25].\par
\smallskip
\indent
The classical action 
$I_{B}$ 
and loop terms 
$A_{0}{\,},
A_{1}{\,},
A_{2}{\,},
{\,}\ldots{\,}$ 
depend in principle on the boundary data.  
In the case of matter coupled to Einstein gravity, 
each of 
${\,}I_{B}{\,},
A_{0}{\,},
A_{1}{\,},
{\,}\ldots$ 
will also obey differential constraints 
connected with the local coordinate invariance of the theory 
and with any other local invariances such as gauge invariance 
(if appropriate) 
[12,26].
This follows from the Dirac canonical approach to the quantisation of theories 
with local or gauge--like invariances [12,26];
the
'differential'
Dirac approach is dual to the 
'integral'
Feynman approach.
The most striking results of the Dirac approach 
arise when the invariance properties include local supersymmetry [12,22,23].
For the present boundary--value problem,
the Dirac approach is much more powerful than the path--integral approach.
Following the Dirac approach.
and given local supersymmetry,
the semi--classical expansion (2.1) simplifies.  
For example, 
in 
$N=1$ 
supergravity, 
one has, 
from the Dirac constrained--quantisation approach [12,24]:
$${\rm Amplitude}
{\;}{\,}
=
{\;}{\,}
A_{0}{\,}\exp\bigl(-{\,}I_{B}/{\hbar}\bigr),
\eqno(2.2)$$
\noindent
apart from factors which are delta--functionals 
of the classical supersymmetry constraints [12,24] on the bounding surfaces.
In this theory, 
the one--loop factor 
$A_{0}$ 
is in fact a constant.  
In Eq.(2.2), 
${\,}I_{B}$ 
denotes the classical action, 
including both bosonic and fermionic contributions.  
Related semi--classical behaviour holds for 
$N=1$ 
supergravity coupled to gauge--invariant supermatter [22,24,27].\par
\smallskip
\indent
In the case (2.2) of 
$N=1$ 
supergravity, 
the classical action is all that is needed for the quantum calculation.  
A corresponding situation arises with ultra--high--energy collisions, 
whether between black holes [28], 
in particle scattering [29], 
or in string theory [30].\par
\smallskip
\indent
In the asymptotically--flat, 
spatially--${\Bbb R^{3}}$ context appropriate here, 
the purely Riemannian case corresponds, 
in Lorentzian--time language, 
to a time--separation at spatial infinity of the rotated form 
$T
=
-{\,}i{\tau}{\,}$, 
where 
${\tau}{\,}$ 
is the 
(positive) 
imaginary--time separation defined above.  
If the four--dimensional classical bosonic part of the solution is to be real, 
then the bosonic boundary data must be chosen real.  
Following the standard route, 
one should study the 
(now complex) 
amplitude (2.1) or (2.2), 
as a function of the angle
${\,}{\theta}{\,}$
of Eq.(1.1),
where
${\,}{\theta}{\,}$ 
is rotated from 
${\,}{\theta}
=
{\pi}/2{\,}$ 
to
${\,}{\theta}
=
+{\,}{\epsilon}{\;}{\,}({\epsilon}>0)$, 
with
$$T
{\;}
=
{\;}
{\tau}{\,}\exp(-{\,}i{\theta}).
\eqno(2.3)$$
\noindent
Provided that there is a 
(complex) 
classical solution to the Dirichlet problem, 
which varies smoothly with
${\theta}{\,},{\;}({\epsilon}\leq{\theta}\leq{\pi}/2)$,
the expression (2.1) or (2.2) should continue to give the quantum amplitude.  
In particular, 
this would occur if strong ellipticity [13] 
held for the coupled Einstein/bosonic--matter field equations, 
up to gauge.\par
\smallskip
\indent
In the present paper,
we assume that the Lagrangian is indeed invariant under local supersymmetry,
as,
for example,
in gauge--invariant
$N=1$
supergravity [27],
but
(for simplicity)
that only the gravitational and scalar data are present in the boundary data 
and for the classical solution.\par
\medskip
\noindent 
{\bf 3. The approximate 4--dimensional metric}
\medskip
\indent
As above, 
the classical background bosonic fields will here, 
for simplicity, 
be taken to be simply the metric 
$g_{{\mu}{\nu}}$ 
and massless scalar field 
${\phi}{\,}$.  
In other work, 
we study cases in which fields with different spins 
$s
=
{{1}\over{2}}{\,},
{\,}1$ 
or 
${{3}\over{2}}$ 
are included as perturbations 
of a spherically--symmetric background solution [2,3,31].  
The classical solutions 
$(g_{{\mu}{\nu}}{\,},{\phi})$ 
of the coupled Einstein/scalar field equations below are taken to have a 
'background' 
time--dependent spherically--symmetric part 
$({\gamma}_{{\mu}{\nu}}{\,},
{\Phi})$, 
together with a 
'small' 
perturbative part 
$(h_{{\mu}{\nu}}{\,},
{\phi}_{\rm pert})$.  
The perturbative fields 
$h_{{\mu}{\nu}}$ 
and 
${\phi}_{\rm pert}{\,}$, 
which live on the spherically--symmetric background 4--geometry with metric
${\gamma}_{{\mu}{\nu}}{\;}$, 
can, 
as usual, 
be expanded out in terms of sums over tensor 
(spin--2), 
vector 
(spin--1) 
and scalar harmonics [32,33].  
Each harmonic is weighted by a function of the Riemannian time 
and radial coordinates 
$({\tau},r)$.
\par
\smallskip
\indent
The Einstein field equations read
$$G_{{\mu}{\nu}}
{\;}{\,}
\equiv
{\;}{\,}
R_{{\mu}{\nu}}-{{1}\over{2}}R{\,}g_{{\mu}{\nu}}
{\;}{\,}
=
{\;}{\,}
8{\pi}{\,}T_{{\mu}{\nu}}
{\;}{\,},
\eqno(3.1)$$
\noindent
where 
$R_{{\mu}{\nu}}$ 
denotes the Ricci tensor, 
$R$ 
the Ricci scalar and 
$T_{{\mu}{\nu}}$ 
the energy--momentum tensor.  
For a real massless scalar field 
${\phi}{\,}$, 
one has 
$$T_{{\mu}{\nu}}
{\;}{\,} 
=
{\;}{\,}
{\phi},_{\mu}{\phi},_{\nu}{\,}
-{\,}{{1}\over{2}}{\,}g_{{\mu}{\nu}}{\,}
\bigl({\phi},_{\alpha}{\phi},_{\beta}{\,}g^{{\alpha}{\beta}}\bigr).
\eqno(3.2)$$  
\noindent
The gravitational field equations further imply the scalar field equation 
(the Laplace--Beltrami equation [34]):
$${\partial_{\mu}}
\bigl(g^{{1}\over{2}}{\,}g^{{\mu}{\nu}}{\phi},_{\nu}\bigr)
{\;} 
=
{\;}
0
{\,},
\eqno(3.3)$$
\noindent
where 
${\,}g{\,}$ 
denotes 
$\det(g_{{\mu}{\nu}})$, 
and 
(for the moment) 
we assume that the 4--metric 
$g_{{\mu}{\nu}}$ 
is real Riemannian, 
whence 
${\,}g>0{\,}$.
\par
\smallskip
\indent
The corresponding variational principle involves an action functional 
of the form [12]
$$I
{\;}
=
{\,}
-{\,}{{1}\over{16{\pi}}}{\,}
{\int}d^{4}x{\;}g^{{1}\over{2}}{\,}R{\;}
+{\,}{{1}\over{2}}{\,}
{\int}d^{4}x{\;}g^{{1}\over{2}}\bigl(\nabla{\phi}\bigr)^{2}
+{\,}{\rm boundary{\;}contributions}.
\eqno(3.4)$$    
\noindent  
The appropriate boundary terms will be discussed in Sec.5.\par
\smallskip
\indent
In the Riemannian case [35], 
the 
'large' 
or 
'background' 
4--metric can be put in the form:
$$ds^{2}
{\;}
=
{\;}
e^{b}{\,}d{\tau}^{2}
+{\,}e^{a}{\,}dr^{2} 
+{\,}r^{2}{\,}
\bigl(d{\theta}^{2}
+{\,}{\sin}^{2}{\theta}{\;}d{\phi}^{2}\bigr),
\eqno(3.5)$$
\noindent
where
$$b
{\;}
=
{\;}
b({\tau},r)
{\;}{\,};
{\;}{\;}{\;}{\;}
a
{\;}
=
{\;}
a({\tau},r).
\eqno(3.6)$$
\noindent
If the gravitational field were exactly spherically symmetric, 
as in Eq.(3.5), 
and if the scalar field were also spherically symmetric,
of the form 
${\phi}({\tau},r)$, 
then the Riemannian spherically--symmetric scalar 
and Einstein field equations would hold [35].  
The scalar field equation reads:
$${\ddot{\phi}}{\,}
+{\,}e^{{b-a}}{\;}{\phi}^{{\prime}{\prime}}
+{\,}{{1}\over{2}}{\,}
\bigl({\dot a}-{\dot b}\bigr){\,}{\dot{\phi}}{\,} 
+{\,}r^{-1}{\,}e^{{b-a}}{\,}
\bigl(1+e^{a}\bigr){\phi}^{\prime}
{\;}
=
{\;}
0
{\,},
\eqno(3.7)$$
\noindent
where 
$({\dot {~~}})$ 
denotes 
${\partial}(~)/{\partial}{\tau}$ 
and 
$(~~)^{\prime}$ 
denotes 
${\partial}(~)/{\partial}r{\,}$.  
Together with Eq.(3.7), 
a slightly redundant set of gravitational field equations is given by:
$$\eqalignno{a^{\prime}
{\;}
&=
{\,}
-{\,}4{\pi}r{\,}
\bigl(e^{a-b}{\,}{\dot{\phi}}^{2}
-{{\phi}^{\prime}}^{2}\bigr)
+{\,}r^{-1}\bigl(1-e^{a}\bigr),
&(3.8)\cr
b^{\prime}
{\;}
&=
{\,}
-{\,}4{\pi}r{\,}
\bigl(e^{a-b}{\,}{\dot{\phi}}^{2}
-{{\phi}^{\prime}}^{2}\bigr)
-{\,}r^{-1}\bigl(1-e^{a}\bigr),
&(3.9)\cr
\dot{a}
{\;}
&=
{\;}{\,}
8{\pi}r{\,}{\dot{\phi}}{\,}{\phi}^{\prime}{\,},
&(3.10)\cr}$$
$${\ddot a}{\,}
+{\,}e^{b-a}{\,}b^{{\prime}{\prime}}
+{\,}{{1}\over{2}}{\,}
\bigl({\dot a}-{\dot b}\bigr){\,}{\dot a}{\,} 
-{\,}r^{-1}{\,}e^{b-a}{\,}
\bigl(1-e^{a}\bigr)
\bigl(b^{\prime}+2r^{-1}\bigr)
=
{\,}
8{\pi}{\,}\bigl({\dot{\phi}}^{2}
+{\,}e^{b-a}{\,}{{\phi}'}^{2}\bigr).
\eqno(3.11)$$ 
\indent
The metric and classical field equations in Lorentzian signature [36], 
or for certain types of complex metrics, 
can be derived from the above by the formal replacement
$$t
{\;}{\,}
=
{\;}{\,}
{\tau}{\,}e^{-{\,}i{\alpha}}{\,},
\eqno(3.12)$$
\noindent 
where 
${\alpha}$ 
is independent of 4--dimensional position, 
and should be rotated from 
$0$ 
to 
${\pi}/2{\;}$.
\par
\smallskip
\indent
In the bosonic black--hole evaporation problem, 
the classical Riemannian metric and scalar field 
will not be exactly spherically symmetric;
similarly for any non--zero spin--${{1}\over{2}}$ 
and spin--${{3}\over{2}}$ classical 
(odd Grassmann--algebra--valued [12]) 
fermionic solutions in any locally--supersymmetric generalisation [27].  
In particle language, 
rather than the field language mostly used in this paper, 
huge numbers of gravitons and scalar particles 
will continually be given off by the black hole 
(together with any fermions allowed by the model), 
leading effectively to a stochastic distribution, 
in which,
for any given spin 
$s{\,}$, 
the field fluctuates around a spherically--symmetric reference field.\par
\smallskip
\indent
Consider, 
for example, 
gravitational and scalar perturbations 
about a Riemannian spherically--symmetric reference 4--metric 
${\gamma}_{{\mu}{\nu}}$
and reference scalar field 
${\Phi}{\,}$.  
In perturbation theory in general relativity [37], 
one considers a one--parameter 
(or many--parameter)
family of 4--metrics, 
here given by the asymptotic expansion
$$g_{{\mu}{\nu}}(x,{\epsilon})
{\;}{\;}{\,}
\sim
{\;}{\;}{\,}
{\gamma}_{{\mu}{\nu}}(x)
+{\,}\epsilon{\,}h^{(1)}_{~~{\mu}{\nu}}(x)
+{\,}{\epsilon}^{2}{\,}h^{(2)}_{~~{\mu}{\nu}}(x)
+{\,}\ldots
{\;}{\,},
\eqno(3.13)$$
\noindent
where 
$h^{(1)}_{~~{\mu}{\nu}}$ 
is the first--order metric perturbation,
$h^{(2)}_{~~{\mu}{\nu}}$ 
the second--order perturbation, 
etc.
Throughout, 
the superscript 
${}^{(0)}$ 
will refer to the background, 
while 
${}^{(1)}$ 
denotes terms linear in perturbations, 
etc.  
Indices are to be raised and lowered using the background metric 
${\gamma}^{{\mu}{\nu}},
{\gamma}_{{\mu}{\nu}}{\,}$.  
Covariant derivatives with respect to the background geometry 
are denoted either by a semi--colon 
${}_{;{\alpha}}$ 
or equivalently by 
${\nabla}_{\alpha}{\;}$.
\par
\smallskip
\indent
Analogously, 
we split a real massless scalar field 
${\,}{\phi}{\,}$ 
into a spherically--symmetric background piece 
${\Phi}({\tau},r)$ 
and a 
(non--spherical) 
perturbation:
$${\phi}(x,{\epsilon})
{\;}{\;}{\,}
\sim
{\;}{\;}{\,}
{\Phi}({\tau},r){\,}
+{\,}{\epsilon}{\,}{\phi}^{(1)}(x){\,}
+{\,}{\epsilon}^{2}{\,}{\phi}^{(2)}(x){\,}
+{\,}\ldots
{\;}{\,}.
\eqno(3.14)$$
\noindent
The spherically--symmetric background part 
${\Phi}$ 
will be non--zero if the background scalar data 
${\phi}$ 
at early and late Euclidean times 
${\tau}$ 
contain a non--trivial spherically--symmetric component.  
The perturbation fields 
${\phi}^{(1)}(x),
{\,}{\phi}^{(2)}(x),
{\,}\ldots{\,}$ 
will, 
in general, 
contain all non--spherical angular harmonics.  
These fields 
must be chosen such that the entire coupled Einstein/scalar system 
satisfies the classical field equations, 
as well as agreeing with the prescribed small non--spherical perturbations 
in the initial and final data, 
both gravitational and scalar.  
The effective energy--momentum source for the spherically--symmetric part 
${\gamma}_{{\mu}{\nu}}$ 
of the metric includes contributions formed quadratically 
from the non--spherical gravitational and scalar expressions 
$h^{(1)}_{~~{\mu}{\nu}}$ 
and 
${\phi}^{(1)}${~}---{~}see Eqs.(3.27--29) below for further detail.\par
\smallskip
\indent
In the simplest case, 
one can restrict attention 
to the exactly spherically--symmetric Riemannian model of Eqs.(3.5--11).  
The background metric 
${\gamma}_{{\mu}{\nu}}{\,}$ 
corresponds to the metric (3.5,6), 
with respect to suitable coordinates, 
and 
${\Phi}({\tau},r)$
above corresponds to  
${\phi}({\tau},r)$ 
of those equations.  
This Riemannian boundary--value problem, 
involving a system of coupled partial differential equations in two variables 
$({\tau},r)$, 
has been studied numerically in [35] for particular choices of boundary data, 
and is currently being investigated in greater detail [38].\par
\smallskip
\indent
In contrast, 
the Lorentzian--signature version 
of the spherically--symmetric classical Einstein/scalar system 
must be studied as an initial--value evolution problem, 
in order to be well posed [36].  
One conceivable initial profile for the scalar field, 
which has been much studied 
in the Lorentzian--signature numerical problem [39,40], 
is an ingoing 
'Gaussian' 
shell of scalar radiation.  
To define such initial data, 
work in a nearly--flat space--time at very early times 
(large negative Lorentzian time--coordinate 
$t$).  
Define an advanced null coordinate
$$v
{\;}
=
{\;}
t{\,}+{\,}r
{\,}.
\eqno(3.15)$$
\noindent
The incoming 
'Gaussian' 
shell is asymptotically, 
at early times, 
of the form:
$${\Phi}(t,r)
{\;}{\;}{\;}
\sim
{\;}{\;}{\;}
{{f_{0}{\,}v^{k+1}}\over{r}}{\,}
\exp\biggl(-{\biggl(}{{v-v_{0}}\over{\Delta}}{\biggr)}^{d}{\,}\biggr),
\eqno(3.16)$$     
\noindent
where 
$f_{0}{\,},
k{\,},
d{\,},
r_{0}{\,}$ 
and 
${\Delta}$ 
are all positive real parameters.  
The radial extent, 
$L_{0}{\;}$, 
of the 
'Gaussian' 
is given by  
${\,}L_{0}{\;}{\,}{\sim}{\;}{\,}{\Delta}{\,}$. 
The numerical evolution of such initial data provides a model 
of spherical collapse.  
In particular, 
two main qualitatively different r\'egimes of initial data 
can be distinguished.  
First, 
if 
$L_{0}$ 
or 
${\Delta}$ 
is too large, 
then the initial data are 
'diffuse', 
there is little self--interaction, 
and the incoming scalar profiles 
pass more or less straight through each other, 
leaving behind nearly--flat space--time plus small perturbations.  
Second, 
if 
${\,}{\Delta}$ 
(or 
$L_{0}$) 
is less than a certain critical value, 
the interaction is sufficiently non--linear that a black hole forms.\par
\smallskip
\indent
Returning to the Riemannian or the complex case, 
one can expand out the Einstein field equations (3.1,2) in powers of 
${\,}{\epsilon}{\,}$.
At lowest order 
$({\epsilon}^{0})$, 
one has the background Einstein and scalar--field equations 
$$\eqalignno{R^{(0)}_{~~{\mu}{\nu}}
-{\,}{{1}\over{2}}{\,}R^{(0)}{\gamma}_{{\mu}{\nu}}
{\;}
&=
{\;}
8{\pi}{\;}T^{(0)}_{~~{\mu}{\nu}}
{\;}{\,},
&(3.17)\cr
{\gamma}^{{\mu}{\nu}}{\,}{\Phi}_{;{\mu}{\nu}}
{\;}
&=
{\;}
0
{\,},
&(3.18)\cr}$$
\noindent
where 
$R^{(0)}_{~~{\mu}{\nu}}$ 
denotes the Ricci tensor and 
$R^{(0)}$
the Ricci scalar of the background geometry
${\gamma}_{{\mu}{\nu}}{\,}$.
Further,
$$T^{(0)}_{~~{\mu}{\nu}}
{\;}
=
{\;}
{\Phi},_{\mu}{\Phi},_{\nu}
-{\,}{{1}\over{2}}{\,}{\gamma}_{{\mu}{\nu}}{\,}
\bigl({\Phi},_{\alpha}{\Phi},_{\beta}{\gamma}^{{\alpha}{\beta}}\bigr)
\eqno(3.19)$$
\noindent
denotes the background spherically--symmetric energy--momentum tensor.
These field equations are equivalent to Eqs.(3.7--11), 
when coordinates are taken as in Eqs.(3.5,6).\par
\smallskip
\indent
The linearised 
$({\epsilon}^{1})$ 
part of the Einstein equations reads
(see Section 35.13 of [16]):
$$\eqalign{&
{\bar{h}}^{(1)~~~~;{\sigma}}_{~~{\mu}{\nu};{\sigma}}  
-2{\,}
{\bar{h}}^{(1)~~~;{\sigma}}_{~~{\sigma}{(}{\mu}~~;{\nu}{)}} 
-2{\,}R^{(0)}_{~~{\sigma}{\mu}{\nu}{\alpha}}{\,}
{\bar{h}}^{(1){\sigma}{\alpha}} 
-2{\,}R^{(0){\alpha}}_{~~~~{(}{\mu}}{\,}
{\bar{h}}^{(1)}_{~~{\nu}{)}{~{\alpha}}}\cr 
&{\quad}{\;}{\;}
+{\,}{\gamma}_{{\mu}{\nu}}{\,}
\bigl({\bar{h}}^{(1)~~;{\alpha}{\beta}}_{~~{\alpha}{\beta}}
-{\bar{h}}^{(1){\alpha}{\beta}}{\,}
R^{(0)}_{~~{\alpha}{\beta}}\bigr)
+{\,}{\bar{h}}^{(1)}_{~{\mu}{\nu}}{\,}R^{(0)}
{\;}
=
{\,}
-16{\pi}{\;}T^{(1)}_{~~{\mu}{\nu}}{\;}{\,}.\cr}
\eqno(3.20)$$
\noindent
Here, 
${\,}{\bar{h}}^{(1)}_{~~{\mu}{\nu}}$ 
is defined by
$${\bar{h}}^{(1)}_{~~{\mu}{\nu}}
{\;}
=
{\;}
h^{(1)}_{~~{\mu}{\nu}}{\,}
-{\,}{{1}\over{2}}{\,}{\gamma}_{{\mu}{\nu}}{\;}h^{(1)}{\,},
\eqno(3.21)$$
\noindent
where 
$$h^{(1)}
{\;}
=
{\;}
h^{(1)~{\mu}}_{~~{\mu}}
{\;}.
\eqno(3.22)$$
\noindent
Also, 
$R^{(0)}_{~~{\sigma}{\mu}{\nu}{\alpha}}$ 
denotes the Riemann tensor of the background geometry 
${\gamma}_{{\mu}{\nu}}{\,}$.  
Further, 
$T^{(1)}_{~~~{\mu}{\nu}}$
denotes the linearisation or 
$O({\epsilon}^{1})$ 
part of the energy--momentum tensor 
$T_{{\mu}{\nu}}(x,{\epsilon})$.  
Explicitly,
$$\eqalign{&T^{(1)}_{~~~{\mu}{\nu}}
{\;}
=
{\;}
2{\,}\nabla_{(\mu}{\phi}^{(1)}{\,}\nabla_{\nu)}{\Phi}{\,}
-{\,}{\gamma}_{{\mu}{\nu}}{\,}\nabla_{\alpha}{\Phi}{\;}{\,}
\nabla^{\alpha}{\phi}^{(1)}\cr
&{\qquad}{\qquad}{\qquad}
+{\,}{{1}\over{2}}{\,}
\Bigl({\gamma}_{{\mu}{\nu}}{\,}h^{(1){\sigma}{\rho}}
-{\,}h^{(1)}_{~~{\mu}{\nu}}{\,}{\gamma}^{{\sigma}{\rho}}\Bigr){\,}
\nabla_{\sigma}{\Phi}{\;}{\,}\nabla_{\rho}{\Phi}
{\;}.\cr}
\eqno(3.23)$$
\indent
The linearised Einstein equations (3.20--23) are most easily studied in a 
'linearised harmonic gauge' 
[16,41] in which, 
by an infinitesimal coordinate transformation, 
one arranges that
$${\bar{h}}^{(1)~~;{\alpha}}_{~~{\mu}{\alpha}}
{\;}
=
{\;}
0 
\eqno(3.24)$$
\noindent
everywhere.  
Since the gravitational background 
${\gamma}_{{\mu}{\nu}}$ 
is spherically symmetric, 
the linearised Einstein equations (3.20--23) 
can be further decomposed into three independent sets of equations.  
These describe, 
repectively, 
scalar 
(spin--0) 
perturbations associated with matter--density changes 
$(T^{(1)}_{~~{\tau}{\tau}})$, 
vector 
(spin--1) 
perturbations associated with matter--velocity changes 
$(T^{(1)}_{~~{\tau}i})$, 
and gravitational radiation 
(spin--2) 
associated with anisotropic stresses 
$(T^{(1)}_{~~ij})$ 
[42].
These equations and their solutions are described further in [2].\par
\smallskip
\indent
The linearised 
$({\epsilon}^{1})$ 
part of the scalar field equation (3.3) yields
$${\gamma}^{{\mu}{\nu}}{\,}{\phi}^{(1)}_{~~;{\mu}{\nu}}{\,}
-{\,}\Bigl({\bar{h}}^{(1){\mu}{\nu}}{\,}{\Phi},_{\nu}\Bigr)_{;{\mu}}
{\;} 
=
{\;}
0
{\,}.
\eqno(3.25)$$
\noindent 
The linearised Einstein and linearised scalar--field equations (3.20--23,25) 
are coupled.\par
\smallskip
\indent
At order 
${\epsilon}^{2}{\,}$, 
the gravitational field equations give 
(implicitly,
below)
the second--order contribution 
$G^{(2)}_{~~{\mu}{\nu}}$ 
to the Einstein tensor
$G_{{\mu}{\nu}}{\;}$.
Note that 
$G^{(2)}_{~~{\mu}{\nu}}$ 
includes a well--known contribution 
quadratic in the first--order perturbations 
$h^{(1)}_{~~{\mu}{\nu}}{\,}$ 
and their derivatives{~}---{~}see Eq.(35.58b) of [16].
This part represents an effective energy--momentum--stress density 
due to the gravitational perturbations, 
including gravitons.
$G^{(2)}_{~~{\mu}{\nu}}$ 
also contains contributions at quadratic order, 
formed from the background 
${\Phi}$ 
and the linearised 
${\phi}^{(1)}$
or their derivatives,
together with 
${\gamma}_{{\mu}{\nu}}$ 
and 
$h^{(1)}_{~~{\mu}{\nu}}{\;}$.  
These parts represent the 
$O({\epsilon}^{2})$ 
contribution of the scalar--field energy--momentum tensor 
$T_{{\mu}{\nu}}$ 
of Eq.(3.2).\par
\smallskip
\indent
Explicitly, 
one finds, 
after a lengthy calculation [1], 
that the Einstein equations, 
to quadratic order in perturbations, 
read
$$G^{(0)}_{~~{\mu}{\nu}}
{\;}
=
{\;}
8{\pi}{\;}T^{(0)}_{~~{\mu}{\nu}}{\,}
+{\,}8{\pi}{\;}T^{(2)}_{~~{\mu}{\nu}}{\,}
+{\,}8{\pi}{\;}T^{\prime}_{{\mu}{\nu}}{\,}
-{\,}G^{(1)}_{~~{\mu}{\nu}}
{\;}{\,}.
\eqno(3.25)$$
\noindent
Here,
$$\eqalign{&T^{(2)}_{~~{\mu}{\nu}}
{\;} 
=
{\;}
\nabla_{\mu}{\phi}^{(1)}{\;}\nabla_{\nu}{\phi}^{(1)}
-{\,}{{1}\over{2}}{\,}{\gamma}_{{\mu}{\nu}}{\,}
{\gamma}^{{\rho}{\sigma}}{\,}
\nabla_{\rho}{\phi}^{(1)}{\;}\nabla_{\sigma}{\phi}^{(1)}\cr
&{\qquad}{\qquad}{\quad}
+{\,}\bigl({\gamma}_{{\mu}{\nu}}{\,}h^{(1){\sigma}{\rho}}
-{\,}h^{(1)}_{~~{\mu}{\nu}}{\,}{\gamma}^{{\sigma}{\rho}}\bigr){\,}
\nabla_{\sigma}{\Phi}{\;}{\,}\nabla_{\rho}{\phi}^{(1)}\cr
&{\qquad}{\qquad}{\quad}
+{\,}{{1}\over{2}}{\,}
\bigl(h^{(1)}_{~~{\mu}{\nu}}{\;}
h^{(1){\sigma}{\rho}}
-{\,}{\gamma}_{{\mu}{\nu}}{\;}h^{(1){\sigma}{\alpha}}{\;}
h^{(1)~{\rho}}_{~~{\alpha}}\bigr){\,}
\nabla_{\sigma}{\Phi}{\;}\nabla_{\rho}{\Phi}\cr}
\eqno(3.27)$$ 
\noindent
and
$$\eqalign{8{\pi}{\;}T'_{{\mu}{\nu}}
{\,}
&=
{\,}
{{1}\over{4}}{\,}
\bigl({\bar{h}}^{(1){\sigma}{\rho}}_{~~~~~;{\mu}}{\,}
h^{(1)}_{~~{\sigma}{\rho};{\nu}}
-{\,}2{\,}{\bar{h}}^{(1)~~;{\alpha}}_{~~{\alpha}{\sigma}}{\;}
{\bar{h}}^{(1){\sigma}}_{~~~({\mu};{\nu})}\bigr)
-{\,}{{1}\over{2}}{\;}
{\bar{h}}^{(1){\sigma}}_{~~~~(\mu}{\,}
R^{(0)}_{~~{\nu}){\rho}{\sigma}{\alpha}}{\;}
{\bar{h}}^{(1){\alpha}{\rho}}\cr
&{\;}{\;}{\;}{\;}
+{\,}{{1}\over{2}}{\;}
{\bar{h}}^{(1)}_{~~{\sigma}({\mu}}{\,}
R^{(0)}_{~~{\nu}){\alpha}}{\,}{\bar{h}}^{(1){\alpha}{\sigma}}
-{\,}{{1}\over{2}}{\;}
h^{(1){\sigma}}_{~~~(\mu}{\,}
{\bar{h}}^{(1)}_{~~{\nu}){\sigma}}{\,}R^{(0)}
-{\,}8{\pi}{\;}
T_{~~{\sigma}({\mu}}^{(1)}{\,}
{\bar{h}}^{(1)~~{\rho}}_{~~{\nu})}\cr
&{\;}{\;}{\;}{\;}
-{\,}4{\pi}{\,}
{\gamma}_{{\mu}{\nu}}{\,}
\bigl(2{\,}{\bar{h}}^{(1){\sigma}{\rho}}{\;}
\nabla_{\sigma}{\phi}^{(1)}{\,}\nabla_{\rho}{\Phi}{\,}
+{\,}{\phi}^{(1)}{\,}\nabla_{\sigma}\nabla^{\sigma}{\phi}^{(1)}
-{\,}{\bar{h}}^{(1){\sigma}{\rho}}{\;}
h^{(1)~{{\beta}}}_{~~{\sigma}}{\,}
\nabla_{\rho}{\Phi}{\,}\nabla_{\beta}{\Phi}\bigr)\cr
&{\;}{\;}{\;}{\;}
+{\,}C^{\sigma}_{~{\mu}{\nu};{\sigma}}
{\;},\cr}
\eqno(3.28)$$
\noindent
where the explicit form of 
$C^{{\sigma}}_{~{\mu}{\nu}}$ 
will not be used here.  
Also, 
$G^{(1)}_{~~{\mu}{\nu}}$
is defined implicitly by Eq.(3.20).
The above expressions 
are needed particularly in studying the Vaidya metric [19], 
which, 
as shown in [21],
describes approximately the late--time region of the geometry 
following gravitational collapse to a black hole, 
containing a nearly--steady outgoing flux of radiation.  
The Einstein field equations, 
averaged over small regions, 
give the contribution of massless scalar particles, 
gravitons, 
etc., 
to the nearly--isotropic flux.\par
\smallskip
\indent
Physically, 
for the Riemannian or complex boundary--value problem discussed in Secs.1 and 
$2{\,}$, 
the 
$O({\epsilon})$ 
perturbations in the 4--metric 
$g_{{\mu}{\nu}}$ 
and scalar field 
${\phi}{\,}$, 
relative to the spherically--symmetric background solution 
$({\gamma}_{{\mu}{\nu}}{\,},
{\Phi})$,
should arise classically from 
$O({\epsilon})$ 
perturbations away from spherical symmetry in the boundary data 
${\,}g_{ij}{\,},
{\Phi}$ 
(or 
${\partial{\Phi}}/{\partial{n}}$) 
at the initial and final surface.
Moreover, 
as mentioned earlier, 
provided that the perturbed boundary data contain numerous high harmonics, 
the 4--dimensional perturbations in the interior 
would be expected effectively to have a stochastic nature.  
When averaged over a number of wavelengths, 
the effective perturbative energy--momentum tensor above, 
$T^{EFF}_{~~~{\mu}{\nu}}{\;}$, 
will yield a spherically--symmetric smoothed--out quantity 
$<T^{EFF}_{~~~{\mu}{\nu}}>{\,}$ 
[43,44].  
In a locally--supersymmetric version of this theory, 
the energy--momentum tensor due to the spin--${{1}\over{2}}$ 
and spin--${{3}\over{2}}$ fields will also contribute to 
$<T^{EFF}_{~~~{\mu}{\nu}}>$.  
In particular,
this averaged form of
$T^{EFF}_{~~~{\mu}{\nu}}$ 
will account for the gradual loss of mass by radiation of a black hole 
in the nearly--Lorentzian sector 
(that is, 
in the case of a time--interval at infinity of the form 
$T{\,}
=
{\,}{\tau}{\,}\exp(-{\,}i{\theta})$, 
where 
${\,}{\tau}{\,}$ 
is large and positive, 
${\,}{\theta}{\,}
=
{\,}{\epsilon}{\,}$ 
is small and positive).  
Although 
$<T^{EFF}_{~~~{\mu}{\nu}}>$ 
is small 
(being of order 
${\epsilon}^{2}$), 
its effects on the black--hole geometry, 
including those on the mass, 
will build up in a secular fashion, 
over a time--scale of order 
${\,}O({\epsilon}^{-2})$.  
Such secular behaviour appears often in perturbation problems 
[45,46]{~}---{~}for example, 
in the familiar treatment of the perihelion precession 
of nearly--circular orbits in the Schwarzschild geometry [47].  
In our boundary--value problem, 
whether regarded as classical or quantum, 
the initial boundary data will be spread over a 
'background' 
extent of 
$O(1)$ 
in the radial coordinate 
$r$ 
on the initial surface 
${\Sigma}_{I}{\;}$.  
But, 
corresponding to the
$O({\epsilon}^{-2})$ 
time--scale for the black hole to radiate, 
the final data on 
${\Sigma}_{F}$ 
will be spread over a radial--coordinate scale of
$O({\epsilon}^{2})$.  
Thus, 
even the classical boundary--value problem here is an example 
of singular perturbation theory [45,46].\par
\smallskip
\indent
The standard treatment of high--frequency averaging in general relativity 
was given by Brill and Hartle [43] and by Isaacson [44].
Let 
$<\;>$ 
denote an average over a time 
$T_{0}{\,}$ 
much longer than typical wave periods, 
together with a spatial average over several wavelengths
${\bar{\lambda}}{\;}$.  
Then:
$$\eqalignno{<g_{{\mu}{\nu}}>
{\;}
&=
{\;}
{\gamma}_{{\mu}{\nu}}
{\;}{\;},
{\qquad}{\qquad}{\;}{\;}
<\phi>
{\;}
=
{\;}
{\Phi}
{\;}{\;},
&(3.29)\cr
<{\phi}^{(1)}>
{\;}
&=
{\;}
0
{\;}{\;},
{\qquad}{\qquad}
<h^{(1)}_{~~{\mu}{\nu}}>
{\;}
=
{\;}
0
{\;}{\;},
&(3.30)\cr
<{\partial}_{\sigma}h^{(1)}_{~~{\mu}{\nu}}>
{\;}
&=
{\;}
0
{\;}{\;},
{\qquad}
<{\partial}_{\sigma}{\partial}_{\rho}
h^{(1)}_{~~{\mu}{\nu}}>
{\;} 
=
{\;}
0
{\;}{\;}.
&(3.31)\cr}$$
\noindent
Indeed, 
$$<C^{(0)}>
{\,}
=
{\;}
C^{(0)}
{\,},
\eqno(3.32)$$
\noindent
for any background quantity 
$C^{(0)}{\,}$.  
Rules for manipulating these averages in the high--frequency aproximation 
are set out in [44].
Under integrals, 
the average of total divergences can be neglected.
For example,
$$<h^{(1){\alpha}~;{\beta}}_{~~~~{\mu}}{\,}
h^{(1)}_{~~{\beta}{\nu};{\alpha}}>
{\,} 
=
{\,}
-{\,}<h^{(1){\alpha}~;{\beta}}_{~~~~{\mu}~~;{\alpha}}{\,}
h^{(1)}_{~~{\beta}{\nu}}>.
\eqno(3.33)$$
\noindent
Further, 
covariant derivatives commute for high--frequency waves.  
The rules (3.29--33) imply that
$$\eqalignno{<T^{(1)}_{~~{\mu}{\nu}}>
{\,}
&=
{\;}
0
{\,},
&(3.34)\cr
<T^{(2)}_{~~{\mu}{\nu}}>
{\,} 
&=
{\,}
<{\nabla}_{(\mu}{\phi}^{(1)}{\,}{\nabla}_{\nu)}{\phi}^{(1)}
-{\,}{{1}\over{2}}{\;}
{\gamma}_{{\mu}{\nu}}{\;}
\nabla_{\alpha}{\phi}^{(1)}{\;}\nabla^{\alpha}{\phi}^{(1)}>.
&(3.35)\cr}$$
\indent
We can now rewrite the background field equations (3.7--11) in a form 
smoothed out by averaging over a number of wavelengths 
of the scalar and gravitational perturbations [43,44].  
The equation which includes the quadratic--order contribution 
of the perturbations as a source for the background geometry reads:
$$G^{(0)}_{~~{\mu}{\nu}}(\gamma)
{\;}
=
{\;}
8{\pi}{\;}T^{(0)}_{~~{\mu}{\nu}} 
+{\,}8{\pi}{\;}{\epsilon}^{2}{\,}
\bigl(<T^{(2)}_{~~{\mu}{\nu}}>
+<T^{\prime}_{{\mu}{\nu}}>\bigr).
\eqno(3.37)$$
\noindent
The terms in this equation vary over length--scales
$\gg{\bar{\lambda}}{\;}$.  
The 
'source equation' 
for 
$h^{(2)}_{~~{\mu}{\nu}}{\;}$, 
analogous to Eq.(3.20) for 
$h^{(1)}_{~~{\mu}{\nu}}{\;}$, 
is
$$G^{(1)}_{~~{\mu}{\nu}}\bigl({\gamma}{\,},h^{(2)}\bigr)
{\,} 
=
{\;}
8{\pi}{\,}
\bigl(T^{\prime}_{{\mu}{\nu}}
-<T^{\prime}_{{\mu}{\nu}}>\bigr)
+{\,}8{\pi}{\,}
\bigl(T^{(2)}_{~~{\mu}{\nu}}
-<T^{(2)}_{~~{\mu}{\nu}}>\bigr).
\eqno(3.38)$$
\noindent
Here, 
the left--hand side 
$G^{(1)}_{~~{\mu}{\nu}}\bigl({\gamma},h^{(2)}\bigr)$ 
denotes the first perturbation of the Einstein tensor 
$G_{{\mu}{\nu}}$ 
about the background metric 
${\gamma}_{{\mu}{\nu}}{\;}$, 
but with its linear argument taken to be 
$h^{(2)}_{~~{\mu}{\nu}}$ 
rather than 
$h^{(1)}_{~~{\mu}{\nu}}{\;}$. 
Thus, 
$-{\,}2{\,}G^{(1)}_{~~{\mu}{\nu}}({\gamma},h^{(1)})$ 
is given by the left--hand side of Eq.(3.20), 
subject to Eqs.(3.21,22).  
Hence, 
the left--hand side of Eq.(3.37) is linear in 
$h^{(2)}_{~~{\mu}{\nu}}$ 
and its derivatives, 
whereas the right--hand side is quadratic in first--order fluctuations.  
By contrast with Eq.(3.36), 
the terms in Eq.(3.37) vary over length--scales of order 
${\bar{\lambda}}{\;}$.
\par
\medskip
\noindent
{\bf 4. Scalar field: harmonic description}
\medskip
\indent
Consider small bosonic perturbations 
${\phi}^{(1)}$ 
and 
$h^{(1)}_{~~{\mu}{\nu}}{\;}$, 
obeying the linearised classical field equations (3.20) and (3.25) 
about a spherically--symmetric classical solution 
$({\Phi},
{\gamma}_{{\mu}{\nu}})$ 
of the Riemannian field equations (3.7--11) for Einstein gravity, 
coupled minimally to a massless scalar field.  
The background spherically--symmetric data for 
${\Phi}$ 
and 
${\,}{\gamma}_{{\mu}{\nu}}{\,}$ 
are posed, 
as in Secs.1,2, 
on the initial and final 3--dimensional boundaries, 
separated at spatial infinity by a 
'Euclidean time--separation' 
${\tau}>0{\,}$.  
Similarly, 
the linearised classical perturbations 
${\phi}^{(1)}$ 
and 
$h^{(1)}_{~~{\mu}{\nu}}$ 
are to be regarded as the solutions to a coupled linear elliptic problem, 
subject to prescribed linearised perturbations 
${\phi}^{(1)}$ 
(say) 
and 
$h^{(1)}_{~~ij}$ 
on the initial and final boundaries.\par
\smallskip
\indent
Because of the spherical symmetry of the background
$({\Phi},
{\gamma}_{{\mu}{\nu}})$, 
one may expand the Riemannian 4--dimensional perturbation 
${\phi}^{(1)}$ 
in the form
$${\phi}^{(1)}({\tau},r,{\theta},{\phi})
{\;} 
=
{\;}
{{1}\over{r}}{\,}\sum^{\infty}_{{\ell}{\,}={\,}0}{\;}
\sum^{\ell}_{m{\,}={\,}-{\,}{\ell}}
Y_{{\ell}m}({\Omega}){\,}R_{{\ell}m}({\tau},r).
\eqno(4.1)$$
\noindent
Here, 
$Y_{{\ell}m}({\Omega})$ 
denotes the 
$({\ell}{\,},m)$ 
scalar spherical harmonic of [48].\par
\smallskip
\indent
Similarly, 
a generic Riemannian metric perturbation 
$h^{(1)}_{~~{\mu}{\nu}}$
may be expanded out as a sum over tensor, 
vector and scalar 
$({\ell}{\,},
m)$ 
harmonics, 
each weighted by a function of 
${\tau}$ 
and 
$r$ 
[32,33,49--52].  
The amplitudes for photon 
(spin--1)
and graviton 
(spin--2) 
emission following black--hole collapse are treated in [2];
there, 
further details of the spin--1 and spin--2 harmonics are given.  
But note that, 
because of the coupled nature of the linearised field equations (3.20,25) for 
${\phi}^{(1)}$ 
and 
$h^{(1)}_{~~{\mu}{\nu}}{\;}$, 
the resulting linear field equations for 
$R_{{\ell}m}({\tau},r)$ 
of Eq.(4.1) and its gravitational analogues 
will also be coupled in the strong--field 
'collapse' 
region of the 
(Riemannian)
`space--time'.\par
\smallskip
\indent
The boundary conditions on the radial functions 
$R_{{\ell}m}({\tau},r)$ 
as 
${\,}r\longrightarrow{0}{\,}$ 
follow from the regularity there of the whole Riemannian solution, 
consisting of 
${\phi}$ 
and the 4--metric 
$g_{{\mu}{\nu}}$ 
(but viewed in 
'nearly--Cartesian coordinates' 
near 
$r=0$). 
This regularity of the solution in turn 
follows since the coupled field equations are 
'elliptic {\it modulo} gauge'.  
For simplicity, 
the boundary data, 
on both the initial and the final 3--surface, 
should be chosen to be suitably regular or smooth over 
${\Bbb R}^{3}$, 
in addition to being asymptotically flat.  
Even when one takes a complex Lorentzian time--separation--at--infinity
$$T
{\;}
=
{\;}
{\tau}{\,}\exp(-{\,}i{\theta}),
\eqno(4.2)$$
\noindent
as in Eq.(2.3), 
with 
${\,}0<{\theta}\leq{\pi}/2{\;}$, 
one expects that the field equations 
(up to gauge) 
will be strongly elliptic [13],
whence all classical fields must be analytic in the interior 
of the large cylindrical boundary formed by the initial and final surfaces,
together with a surface at large 
$r{\,}$.
\par
\smallskip
\indent
Suppose that the boundary conditions on the final surface 
are taken to describe very weak and diffuse scalar and gravitational fields,
to be regarded as perturbations of flat 3--space 
${\Bbb E}^{3}{\,}$. 
(One also requires that the 
$ADM$ 
mass of the final intrinsic boundary 3--metric 
$g_{ij}{\;}$, 
as computed from the 
$1/r$
part of the fall--off of 
$g_{ij}$ 
to the flat metric 
${\delta}_{ij}{\,}$ 
[16,17,53], 
should be the same as the 
$ADM$ 
mass of the initial surface.
This will be discussed further in Sec.5 below.)  
Physically, 
such weak and diffuse final boundary data 
may be imagined to be a possible late--time remnant of gravitational collapse, 
namely, 
a snap--shot of a large number of scalar particles and gravitons, 
as they make their way out to infinity.  
Near the final surface, 
the coupling in Eqs.(3.20,25) between the linearised perturbations 
${\,}{\phi}^{(1)}$ 
and 
$h^{(1)}_{~~{\mu}{\nu}}$ 
will almost have disappeared.  
The perturbed scalar field equation at late times is simply
$${\nabla}^{\mu}{\nabla}_{\mu}{\phi}^{(1)}
{\;}
=
{\;}
0
{\,},
\eqno(4.3)$$
\noindent
with respect to the spherically--symmetric background geometry
${\gamma}_{{\mu}{\nu}}{\;}$.
\par
\smallskip
\indent
Making the mode decomposition (4.1) of 
${\phi}^{(1)}$, 
one obtains the 
$({\ell}{\,},m)$ 
mode equation
$$\bigl(e^{(b-a)/2}{\,}\partial_{r}\bigr)^{2}R_{{\ell}m} 
+\bigl(\partial_{\tau}\bigr)^{2}R_{{\ell}m} 
+{\,}{{1}\over{2}}{\,}\bigl(\partial_{\tau}(a-b)\bigr)
\bigl(\partial_{\tau}R_{{\ell}m}\bigr)
-V_{\ell}({\tau},r){\,}R_{{\ell}m}
{\;}
=
{\;}
0
{\,}.
\eqno(4.4)$$
\noindent
Here,
the potential
$V_{\ell}({\tau},r)$
is given by
$$V_{\ell}({\tau},r)
{\;}
=
{\;}
{{e^{b({\tau},r)}\over{r^{2}}}}{\,}
{\Bigl({\ell}({\ell}+1)+{\,}{{2m({\tau},r)}\over{r}}\Bigr)},
\eqno(4.5)$$
\noindent
and 
$m({\tau},r)$ 
is defined by 
$$\exp\bigl(-{\,}a({\tau},r)\bigr)
{\;}
=
{\;}
1{\,}-{\,}{{2m({\tau},r)}\over{r}}
{\;}{\,}.
\eqno(4.6)$$
\noindent
In an exact Schwarzschild solution with no scalar field, 
one would have 
${\,}e^{b}
=
e^{-a}
=
1-(2M/r)$, 
with 
$M$ 
the Schwarzschild mass; 
in that case, 
${\,}m({\tau},r)$ 
would be identically 
$M{\,}$.  
The potential 
$V_{\ell}({\tau},r)$ 
of Eq.(4.5) 
generalises the well--known massless--scalar effective potential 
in the exact Schwarzschild geometry [16], 
which vanishes at the event horizon 
$\{r=2M\}$ 
and at spatial infinity, 
and has a peak near 
$\{r=3M\}$.
\par
\smallskip
\indent
The definition (4.6) of 
$m({\tau},r)$ 
is also consistent with the usual description 
of the Lorentzian--signature Vaidya metric [19,20].  
In terms of a null coordinate 
$u{\,}$ 
and an intrinsic radial coordinate 
$r{\,}$,
the Vaidya metric reads
$$ds^{2}
{\;}
=
{\,}
-2{\,}du{\,}dr{\,}
-\Bigl(1-{{2m(u)}\over{r}}\Bigr){\,}du^{2} 
+{\,}r^{2}{\,}
\bigl(d{\theta}^{2}+{\,}\sin^{2}{\theta}{\;}d{\phi}^{2}\bigr).
\eqno(4.7)$$
\noindent
Here, 
$m(u)$ 
is a monotonic--decreasing smooth function of 
$u{\,}$, 
corresponding to a suitable spherically--symmetric outflow of null particles, 
for example by taking the energy--momentum tensor of a black hole evaporating 
{\it via} 
emission of scalar particles at the speed of light.  
The Vaidya metric 
has been used often to give an approximate gravitational background 
for black--hole evaporation at late times [54--56].  
In connection with our present work, 
the Vaidya geometry has been treated in [21].\par
\smallskip
\indent
There is, 
of course, 
an analogous decoupled harmonic decomposition,
valid near the final surface, 
for the weak gravitational--wave perturbations 
about the spherically--symmetric background{~}---{~}again described in [2].  
For simplicity of exposition, 
we shall here restrict attention to weak--field final configurations 
for spin 0 
(scalar), 
and calculate their quantum amplitudes, 
on the further assumption that the final 3--metric 
$h_{ijF}$ 
is exactly spherically symmetric 
(in addition to the assumed spherical symmetry of the initial data 
${\,}{\phi}_{I}{\,}$ 
and 
$h_{ijI}$).  
Once the methods are established in the simplest spin--$0$ case, 
generalisation to the case of higher--spin fields 
becomes relatively straightforward.\par
\medskip
\noindent
{\bf 5. The classical action}
\medskip
\indent
Consider, 
for definiteness, 
an asymptotically--flat Lorentzian--signature classical solution 
$(g_{{\mu}{\nu}}{\,},
{\phi})$ 
of the coupled Einstein/massless--scalar field equations, 
between an initial hypersurface 
${\Sigma}_{I}$ 
and a final hypersurface 
${\Sigma}_{F}{\;}$, 
separated by a Lorentzian proper time
$T$ 
at spatial infinity.  
Write 
$S$ 
for the Lorentzian action functional, 
which corresponds to the Riemannian action functional 
$I$ 
of Eq.(3.4) with suitable boundary contributions [12], 
appropriate to fixing the boundary data
$(h_{ij}{\,},
{\phi})_{I}{\,}$ 
and 
$(h_{ij}{\,},
{\phi})_{F}{\;}$, 
according to 
${\,}iS
=
{\,}-{\,}I{\,}$.   
At the Lorentzian--signature solution above, 
one has [12,57] the classical action
$$\eqalign{&S_{\rm class}\bigl[(h_{ij}{\,},{\phi})_{I}{\,};
(h_{ij}{\,},{\phi})_{F}{\,};
T\bigr]\cr
&{\quad}{\;}{\;}
=
{\;}
{{1}\over{32{\pi}}}{\,}
\biggl(\int_{{\Sigma}_{F}}-\int_{{\Sigma}_{I}}\biggr){\,}
d^{3}x{\;}{\,}{\pi}^{ij}{\;}h_{ij}{\,}
+{\,}{{1}\over{2}}{\,}
\biggl(\int_{{\Sigma}_{F}}-\int_{{\Sigma}_{I}}\biggr){\,}
d^{3}x{\;}{\,}{\pi}_{\phi}{\;}{\phi}{\,}
-{\,}MT
{\,}.\cr}
\eqno(5.1)$$
\noindent
Here, 
${\pi}^{ij}
=
{\pi}^{ji}$ 
is 
$16{\pi}$ 
times the Lorentzian momentum conjugate to the 
'coordinate' 
variable 
$h_{ij}$ 
on a space--like hypersurface, 
in a 
$3+1$ 
Hamiltonian decomposition of the Einstein/massless--scalar theory [58].  
Explicitly, 
in terms of the Lorentzian--signature second fundamental form 
$K_{ij}
=
K_{(ij)}$ 
of the hypersurface [12,41], 
${\,}{\pi}^{ij}$ 
is given by
$${\pi}^{ij}
{\;}
=
{\;}
h^{{1}\over{2}}{\,}\bigl(K^{ij}-K{\,}h^{ij}\bigr),
\eqno(5.2)$$
\noindent
where 
$h
=
{\rm det}(h_{ij})$ 
and 
$K
=
h^{ij}K_{ij}{\;}$.  
Further, 
${\pi}_{\phi}$ 
is the Lorentzian momentum conjugate to the 
'coordinate' 
variable 
${\,}{\phi}{\,}$.  
Explicitly,   
$${\pi}_{\phi}
{\;}
=
{\;}
h^{{1}\over{2}}{\;}n^{\mu}\nabla_{\mu}{\phi}
{\;},
\eqno(5.3)$$ 
\noindent
where 
${\,}n^{\mu}$ 
denotes the 
(Lorentzian--signature) 
future--directed unit time--like vector normal to the hypersurface.\par
\smallskip
\indent
Suppose instead that one has a complex or a Riemannian solution
$(g_{{\mu}{\nu}}{\,},{\phi})$ 
between asymptotically--flat boundary data
$(h_{ij}{\,},{\phi})_{I}{\,}$ 
and 
$(h_{ij}{\,},{\phi})_{F}{\,}$ 
on initial and final hypersurfaces 
${\Sigma}_{I}{\,},
{\Sigma}_{F}{\,}$, 
where the time--separation 
$T$
at infinity has the form 
$T
=
{\tau}\exp(-{\,}i{\theta})$, 
as in Eq.(2.3), 
with
${\tau}$ 
positive real, 
$(0<{\theta}\leq{\pi}/2)$.  
As above, 
this is expected to provide the natural arena 
for asymptotically--flat boundary--value problems involving gravitation, 
provided strong ellipticity holds, 
up to gauge.  
For such a solution, 
the Lorentzian--signature classical action 
$S_{\rm class}$ 
is still defined by Eq.(5.1).  
This action will in general be complex, 
although for a real Riemannian solution with 
${\theta}
=
{\pi}/2{\;}$, 
the Riemannian action 
$I_{\rm class}{\;}$, 
defined by 
$I_{\rm class}
=
{\,}-{\,}iS_{\rm class}{\;}$, 
is real.
The boundary contribution at spatial infinity to the Riemannian action 
functional 
$I$ 
of Eq.(5.1) is 
$M{\tau}$ 
[12].  
The boundary contributions to the functional 
$I{\,}$, 
due to the presence of the boundaries 
${\Sigma}_{I}$ 
and 
${\Sigma}_{F}$ 
with specified data 
$(h_{ij}{\,},
{\phi})_{I}$ 
and 
$(h_{ij}{\,},
{\phi})_{F}{\;}$, 
are 
$$I_{I}{\,}+{\,}I_{F}
{\;}
=
{\;}
{{1}\over{32{\pi}}}{\,}
\biggl(\int_{{\Sigma}_{I}}-\int_{{\Sigma}_{F}}\biggr){\,}
d^{3}x{\;}{\,}{}_{e}{\pi}^{ij}{\;}h_{ij}{\,}
+{\,}{{1}\over{2}}{\;}
\biggl(\int_{{\Sigma}_{F}}-\int_{{\Sigma}_{I}}\biggr){\,}
d^{3}x{\;}{\,}{}_{e}{\pi}_{\phi}{\;}{\phi}
{\;}.
\eqno(5.4)$$
\noindent
Here,
$${}_e{\pi}^{ij}
{\;}
=
{\;}
h^{{1}\over{2}}{\,}
\Bigl({}_{e}K^{ij}{\,}-{\,}{}_{e}K{\;}h^{ij}\Bigr)
\eqno(5.5)$$
\noindent
is given by the same formula as 
${\pi}^{ij}$ 
in Eq.(5.2), 
except that 
$K_{ij}$ 
has been replaced by the 
'Euclidean' 
second fundamental form
${}_{e}K_{ij}{\;}$, 
as defined and used in Eqs.(2.6.23,24) of [12].  
In particular, 
$${}_{e}K_{ij}
{\;}
=
{\,}
-i{\,}K_{ij}
{\;}{\,}.
\eqno(5.6)$$
\noindent
Similarly, 
the scalar--momentum variable 
${\pi}_{\phi}$ 
of Eq.(5.3) has been replaced by its 
'Euclidean' 
version 
${}_{e}{\pi}_{\phi}{\;}$, 
defined by 
$${}_{e}{\pi}_{\phi}
{\;}
=
{\;}
h^{{1}\over{2}}{\;}{}_{e}n^{\mu}\nabla_{\mu}{\phi}
{\;}{\,},
\eqno(5.7)$$
\noindent
where [12]
$${}_{e}n^{\mu}
{\;}
=
{\,}
-i{\,}n^{\mu}
\eqno(5.8)$$
\noindent
denotes the unit future--directed Riemannian normal.\par
\smallskip
\indent
The quantity 
$M$ 
in Eq.(5.1) is the 
$ADM$ 
mass of the 
'space--time', 
as measured near spatial infinity from the 
$1/r{\,}$ 
part of the fall--off of the intrinsic spatial metric
$h_{ij}{\,}$ 
on 
${\Sigma}_{I}$ 
and 
${\Sigma}_{F}$ 
[16,53].  
As mentioned in Sec.1, 
it is essential, 
for a well--posed asymptotically--flat boundary--value problem, 
that the intrinsic metrics 
$h_{ijI}$ 
and 
$h_{ijF}$ 
be chosen to have the same value of 
$M{\,}$.  
Otherwise, 
if 
${\,}{M}_{I}\neq{M}_{F}{\;}$, 
then any classical infilling 
'space--time' 
will have
${\Sigma}_{I}$ 
and 
${\Sigma}_{F}$ 
badly embedded near spatial infinity, 
and the entire 4--metric 
$g_{{\mu}{\nu}}$ 
will not fall off to flatness at the standard
$1/r$
rate, 
as 
${\,}r\longrightarrow{\infty}{\,}$ 
[17].\par
\smallskip
\indent
In applications to particle emission, 
including the case of nearly--spherical collapse to a black hole, 
we naturally make use of the perturbative splitting
$$\eqalignno{g_{{\mu}{\nu}}
{\;}{\;}{\;}{\,} 
&{\sim}
{\;}{\;}{\;}{\,}
{\gamma}_{{\mu}{\nu}}{\,}
+{\,}h^{(1)}_{~~{\mu}{\nu}}{\,}
+{\,}h^{(2)}_{~~{\mu}{\nu}}{\,}
+{\,}\ldots
{\;}{\,},
&(5.9)\cr
{\phi}
{\;}{\;}{\;}{\,} 
&{\sim}
{\;}{\;}{\;}{\,}
{\Phi}{\,}
+{\,}{\phi}^{(1)}{\,}
+{\,}{\phi}^{(2)}{\,}
+{\,}\ldots
{\;}{\;}. 
&(5.10)\cr}$$
\noindent
Here, 
the spherically--symmetric 
'background' 
solution 
$({\gamma}_{{\mu}{\nu}}{\,},
{\Phi})$ 
obeys the coupled Einstein /massless--scalar classical field equations, 
as does the full classical solution 
$(g_{{\mu}{\nu}}{\,},
{\phi})$.  
(The formal device of including a small parameter 
${\epsilon}$ 
has been relaxed here;  
we now set 
${\,}{\epsilon}=1$.)
\par
\smallskip
\indent
The linearised fields 
$h^{(1)}_{~~{\mu}{\nu}}$ 
and 
${\phi}^{(1)}$ 
may be decomposed into sums of appropriate angular harmonics, 
labelled by quantum numbers 
$({\ell}{\,},
m)$, 
as in Sec.4, 
and without loss of generality 
it may be assumed that any spherically--symmetric 
${\ell}=0$ 
linear--order perturbation modes 
have been absorbed into the spherically--symmetric background 
$({\gamma}_{{\mu}{\nu}}{\,},
{\Phi})$.  
Then 
(say) 
the Lorentzian classical action 
$S_{\rm class}$ 
of Eq.(5.1) may be split as
$$S_{\rm class}
{\;} 
=
{\;}
S^{(0)}_{\rm class}{\,}
+{\,}S^{(2)}_{\rm class}{\,}
+{\,}S^{(3)}_{\rm class}{\,}
+{\,}\ldots
{\;}{\;}.
\eqno(5.11)$$
\noindent
Here, 
$S^{(0)}_{\rm class}$ 
is the background action, 
given by Eq.(5.1), 
but evaluated for the spherically--symmetric solution
$({\gamma}_{{\mu}{\nu}}{\,},
{\Phi})$.  
The mass 
$M$ 
appearing in 
$S^{(0)}_{\rm class}$ 
will be that determined from 
${\gamma}_{ijI}$ 
or 
${\gamma}_{ijF}{\;}$.  
The next term is 
$S^{(2)}_{\rm class}{\;}$, 
formed quadratically from the linear--order perturbations; 
the linear--order term
$S^{(1)}_{\rm class}$ 
is zero, 
because one is perturbing around a classical solution.  
In an obvious notation, 
one has 
$$S^{(2)}_{\rm class}
{\;} 
=
{\;}
{{1}\over{32{\pi}}}{\;}
\biggl(\int_{{\Sigma}_{F}}-\int_{{\Sigma}_{I}}\biggr){\,}
d^{3}x{\;}{\,}{\pi}^{(1)ij}{\;}h^{(1)}_{~~ij}{\,}
+{\,}{{1}\over{2}}{\;}
\biggl(\int_{{\Sigma}_{F}}-\int_{{\Sigma}_{I}}\biggr){\,}
d^{3}x{\;}{\,}{\pi}^{(1)}_{\phi}{\;}{\phi}^{(1)}
{\,}.
\eqno(5.12)$$ 
\noindent
Thus, 
$S^{(2)}_{\rm class}$ 
is contructed only from quantities on the boundaries 
${\Sigma}_{I}$ 
and 
${\Sigma}_{F}{\;}$.
Note that there is no contribution to the second--order expression
$S^{(2)}_{\rm class}$ 
from the 
$-{\,}MT$ 
term in Eq.(5.1), 
again because of the above definitions.\par
\smallskip
\indent
The expression (5.1) for 
${\,}S_{\rm class}[(h_{ij}{\,},{\phi})_{I}{\;};
{\,}(h_{ij}{\,},{\phi})_{F}{\;};
{\,}T]$, 
together with the asymptotic series (5.9) for the classical action 
and the expression (5.10) for 
$S^{(2)}_{\rm class}$ 
formed from the linearised perturbations, 
will be basic in calculations concerning quantum amplitudes 
in subsequent work.\par
\medskip
\noindent
{\bf 6. Adiabatic radial functions} 
\medskip
\indent
We return to the evolution of linearised scalar--field perturbations 
${\phi}^{(1)}$,  
following the mode sum 
(angular decomposition) 
of Eq.(4.1).  
For the quantum amplitudes of interest, 
we must compute expressions of the form 
$${\rm Amplitude}
{\;} 
=
{\;}
{\rm const.}{\,}{\times}{\,}
\exp\Bigl\{i{\,}S_{\rm class}
\bigl[\bigl(h_{ij}{\,},{\phi}\bigr)_{I}{\;};
{\,}\bigl(h_{ij}{\,},{\phi}\bigr)_{F}{\;};
{\,}T\bigr]\Bigr\},
\eqno(6.1)$$  
\noindent
where, 
equivalently, 
$iS_{\rm class}
=
{\,}-{\,}I_{\rm class}{\;}$.  
As above, 
the time--interval 
$T{\,}$, 
measured at spatial infinity, 
must be of the form
${\,}T
=
{\mid}T{\mid}\exp(-{\,}i{\theta}),  
{\;}(0<{\theta}\leq{\pi}/2)$, 
for a classical solution to exist for the boundary--value problem.  
The classical action 
$S_{\rm class}$ 
is given in Eq.(5.1) in terms of integrals taken over the boundaries 
${\Sigma}_{I}$
and 
${\Sigma}_{F}{\;}$, 
subject to the classical field equations.\par
\smallskip
\indent
Consider the amplitude corresponding to weak--field non--spherical data 
$(h^{(1)}_{~ij})_{F}$ 
and 
$({\phi}^{(1)})_{F}$ 
on the final surface, 
given at lowest order by  
$\exp\bigl(iS^{(2)}_{\rm class}\bigr)$.  
For simplicity, 
take the initial data to be exactly spherically symmetric, 
namely, 
$({\gamma}_{ij}{\,},
{\Phi})_{I}{\;}$.  
Equivalently, 
$$h_{~~ijI}^{(1)}
{\;}
=
{\;}
0
{\;}{\,};
{\;}{\;}{\;}{\,}
{\phi}^{(1)}_{~~I}
{\;}
=
{\;}
0
{\,}.
\eqno(6.2)$$   
\noindent
The amplitude 
$\exp(iS_{\rm class})$ 
will then depend only on the contributions at the final surface 
${\Sigma}_{F}$ 
in Eq.(5.12) 
[which themselves depend on
$(h^{(1)}_{~~ijF}{\;},
{\phi}^{(1)}_{~F}{\;};
{\,}T)$].  
As a practical matter, 
one could easily put non--zero 
$(h^{(1)}_{~~ij}{\;},
{\,}{\phi}^{(1)})_{I}{\,}$ 
back into the calculations that follow.  
Physically, 
the analogous step of 
'turning back on the early--time perturbations' 
corresponds, 
in 
'particle language' 
rather than in the 
'field language' 
being used in this paper, 
to the inclusion of extra particles in the in--states, 
together with the original spherical collapsing matter, 
and asking for the late--time consequences.  
This was first carried out by Wald [59].\par
\smallskip
\indent
In this paper, 
we concentrate only on the scalar--field contribution to the quantum amplitude 
$\exp(iS_{\rm class})$.  
That is, 
we compute
$$S^{(2)}_{\rm class{\,},{\,}scalar}
{\;} 
=
{\;}
{{1}\over{2}}{\,}
\int_{{\Sigma}_{F}}d^{3}x{\;}{\,}
{\pi}^{(1)}_{\phi}{\;}{\phi}^{(1)},
\eqno(6.3)$$
\noindent
where the linearised perturbations 
$(h^{(1)}_{~~{\mu}{\nu}}{\;},
{\phi}^{(1)})$
obey the linearised field equations (3.20--22,25) 
about the spherically--symmetric background 
$({\gamma}_{{\mu}{\nu}}{\,},
{\Phi})$.  
Here,
$(h^{(1)}_{~~{\mu}{\nu}}{\,},
{\phi}^{(1)})$ 
must agree with the prescribed final data 
$(h^{(1)}_{~~ij}{\;},
{\phi}^{(1)})_{F}$ 
at the final surface 
${\Sigma}_{F}{\;}$, 
and be zero at the initial surface 
${\Sigma}_{I}{\;}$.  
In the Riemannian case, 
with a real Euclidean time--interval 
${\tau}$ 
between 
${\Sigma}_{I}$ 
and 
${\Sigma}_{F}{\;}$, 
or in the case (2.3) of a complex time--interval 
$T
=
{\tau}\exp(-{\,}i{\theta})$ 
between 
${\Sigma}_{I}$ 
and 
${\Sigma}_{F}{\;}$, 
with 
$0<{\theta}<{\pi}/2{\;}$, 
this linear boundary--value problem is expected to be well--posed.  
The other, 
gravitational, 
contribution
$$S^{(2)}_{\rm class{\,},{\,}grav}
{\;} 
=
{\;}
{{1}\over{32{\pi}}}{\,}
\int_{{\Sigma}_{F}}d^{3}x{\;}{\,}
{\pi}^{(1)ij}{\;}h^{(1)}_{~~ij}
\eqno(6.4)$$
\noindent
to 
$S^{(2)}_{\rm class}$ 
in Eq.(5.12) is studied in [2].\par
\smallskip
\indent
Following Sec.4, 
at late times the perturbed scalar--field equation reduces to
$$\nabla^{\mu}\nabla_{\mu}{\phi}^{(1)}
{\,} 
=
{\;}
0
{\,},
\eqno(6.5)$$  
\noindent
with respect to the spherically--symmetric background 
${\gamma}_{{\mu}{\nu}}{\;}$.  
Here, 
in contrast to Eq.(3.5), 
it is more suitable to work with the Lorentzian background 
gravitational field: 
$$ds^{2}
{\;} 
=
{\,}
-{\,}e^{b(t,r)}{\,}dt^{2}{\,}
+{\,}e^{a(t,r)}{\,}dr^{2}{\,}
+{\,}r^{2}{\,}
(d{\theta}^{2}+{\,}\sin^{2}{\theta}{\;}d{\phi}^{2}).
\eqno(6.6)$$   
\noindent
By analogy with Eq.(4.1), 
one makes the mode decomposition with respect to 
'Lorentzian coordinates'  
$(t,r,{\theta},{\phi})$:
$${\phi}^{(1)}(t,r,{\theta},{\phi})
{\;} 
=
{\;}
{{1}\over{r}}{\,}
\sum^{\infty}_{\ell{\,}={\,}0}{\;}
\sum^{\ell}_{m{\,}={\,}-{\,}\ell}
Y_{{\ell}m}(\Omega){\,}R_{{\ell}m}(t,r).
\eqno(6.7)$$
\noindent
As in Eq.(4.4), 
one arrives at the 
$({\ell}{\,},m)$ 
mode equation:
$$\bigl(e^{(b-a)/2}{\,}\partial_{r}\bigr)^{2}R_{{\ell}m}
-\bigl(\partial_{t}\bigr)^{2}R_{{\ell}m}
-{\,}{{1}\over{2}}{\,}\bigl(\partial_{t}\bigl(a-b\bigr)\bigr)
\bigl(\partial_{t}R_{{\ell}m}\bigr)
-V_{\ell}(t,r){\,}R_{{\ell}m}
{\;}
=
{\;}
0
{\,}.
\eqno(6.8)$$
\noindent
Here, 
$V_{\ell}(t,r)$ 
is defined by Eq.(4.5), 
except that,
in all its appearances, 
the argument 
${\tau}$ 
is replaced by 
$t{\,}$.  
Similarly, 
the function 
$m(t,r)$ 
is defined by Eq.(4.6), 
with a corresponding replacement of 
${\tau}$ 
by 
$t{\,}$.
\par
\smallskip
\indent
For high frequencies of oscillation in the nearly--Lorentzian case,
with small angle 
${\theta}$ 
of rotation into the complex, 
it becomes simpler to understand the solutions of the mode equation (6.8).
Consider a solution 
$R_{{\ell}m}(t,r)$ 
of the form
$$R_{{\ell}m}(t,r)
{\;}{\;}{\;}
{\sim}
{\;}{\;}{\;}
\exp(ikt){\;}{\xi}_{k{\ell}m}(t,r),
\eqno(6.9)$$
\noindent
where 
$k{\,}$ 
is a 
'large' 
frequency, 
but where, 
in contrast, 
${\,}{\xi}_{k{\ell}m}(t,r)$ 
varies 
'slowly' 
with respect to 
$t{\,}$.  
In particular, 
we require that, 
near spatial infinity, 
with 
$r\longrightarrow{\infty}{\,}$, 
$R_{{\ell}m}(t,r)$ 
reduces to a flat space--time separated solution, 
in which 
${\xi}_{k{\ell}m}(t,r)$ 
loses its $t$--dependence 
[see Eqs.(6.11,14) below].\par
\smallskip
\indent
Our boundary--value problem is for scalar perturbations 
${\phi}^{(1)}(t,r,{\theta},{\phi})$,
or equivalently for functions 
$R_{{\ell}m}(t,r)$ 
as in Eqs.(6.7,8), 
subject to the initial condition 
${\phi}^{(1)}{\bigl\vert}_{t=0}
=
0$ 
and to prescribed real final data
${\phi}^{(1)}{\bigl\vert}_{t=T}{\;}$.  
Were the propagation simply in flat space--time, 
the solution would be of the form 
$${\phi}^{(1)}
{\,} 
=
{\;}
{{1}\over{r}}{\;}
\sum^{\infty}_{\ell{\,}={\,}0}{\;}
\sum^{\ell}_{m{\,}={\,}-{\,}\ell}{\;}
\int^{\infty}_{-\infty}dk{\;}{\,}a_{k{\ell}m}{\;}
{\xi}_{k{\ell}m}(r){\,}{{\sin(kt)}\over{\sin(kT)}}{\;}
Y_{{\ell}m}(\Omega),
\eqno(6.10)$$
\noindent 
where the 
$\{a_{k{\ell}m}\}$ 
are real coefficients and each function 
${\xi}_{k{\ell}m}(r)$ 
is proportional 
(up to a factor of $r$)
to a spherical Bessel function
$j_{\ell}(kr)$ 
[60].  
In our gravitational--collapse case, 
${\xi}_{k{\ell}m}$ 
becomes a function of 
$t{\,}$ 
as well as of 
$r{\,}$, 
but otherwise the pattern remains:
$${\phi}^{(1)}
{\,}
=
{\;}
{{1}\over{r}}{\,}
\sum^{\infty}_{{\ell}{\,}={\,}0}{\;}
\sum^{\ell}_{m{\,}={\,}-{\,}{\ell}}{\;}
\int^{\infty}_{-\infty}dk{\;}{\,}a_{k{\ell}m}{\;}
{\xi}_{k{\ell}m}(t,r){\;}{{\sin(kt)}\over{\sin(kT)}}{\;}
Y_{{\ell}m}(\Omega).
\eqno(6.11)$$ 
\noindent
Here, 
the 
$\{a_{k{\ell}m}\}$ 
characterise the final data:  
they can be constructed from the given 
${\phi}^{(1)}{\bigl\vert}_{t=T}$ 
by inverting Eq.(6.11).  
The functions 
${\xi}_{k{\ell}m}(t,r)$ 
are defined in the adiabatic or large--${\mid}k{\mid}$ limit, 
as in the previous paragraph, 
{\it via} 
Eq.(6.9), 
where 
$R_{{\ell}m}(t,r)$ 
obeys the mode equation (6.8).\par
\smallskip
\indent
More precisely, 
provided that 
$k$ 
is large, 
in the sense that the adiabatic approximation 
$${\bigl\vert}k{\bigl\vert}
{\;}
\gg
{\;}
{{1}\over{2}}{\,}{\bigl\vert}{\dot a}-{\dot b}{\bigl\vert},
\eqno(6.12)$$
$${\bigl\vert}k{\bigl\vert}
{\;}
\gg
{\;}
{\biggl\arrowvert}
{{{\dot{\xi}}_{k{\ell}m}}\over{{\xi}_{k{\ell}m}}}
{\biggl\arrowvert}, 
{\quad}
k^{2}
{\;} 
\gg
{\;}
{\biggl\arrowvert}
{{{\ddot{\xi}}_{k{\ell}m}}\over{{\xi}_{k{\ell}m}}}
{\biggl\arrowvert}
\eqno(6.13)$$
\noindent
holds, 
the mode equation reduces approximately to 
$$e^{(b-a)/2}{\,}{{\partial}\over{\partial r}}
\Bigl(e^{(b-a)/2}{\,}{{\partial{\xi}_{k{\ell}m}}\over{\partial{r}}}\Bigr) 
+{\,}\bigl(k^{2}-V_{\ell}\bigr){\,}{\xi}_{k{\ell}m}
{\;}
=
{\;}
0
{\,}.
\eqno(6.14)$$
\noindent
Of course, 
the functions 
$e^{(b-a)/2}$ 
and 
$V_{\ell}$ 
do still vary with the time--coordinate 
$t{\,}$, 
but only adiabatically or 
'slowly'.\par
\smallskip
\indent
As described further in [1,5,19], 
the geometry in the radiative region of the space--time 
is expected to be approximated very accurately 
by a spherically--symmetric Vaidya metric [19,20], 
corresponding to a luminosity in the radiated particles 
which varies slowly with time.  
Such a metric can be put in the diagonal form
$$e^{-a}
{\;}
=
{\;}
1{\,}-{\,}{{2m(t,r)}\over{r}}
{\;}{\;};
{\;}{\;}{\;}{\;}{\,}
e^{b}
{\;}
=
{\;}
\biggl({{\dot m}\over{f(m)}}\biggr)^{2}{\,}e^{-a}
{\,},
\eqno(6.15)$$
\noindent
where 
$m(t,r)$ 
is a slowly--varying function, 
with 
${\dot m}
=
(\partial m/\partial{t})$, 
and where the function 
$f(m)$
depends on the details of the radiation.  
Then Eq.(6.12) implies that
$${\bigl\vert}k{\bigl\vert}
{\;}{\,}
\gg
{\;}{\,}
{\Bigl\vert}{{\dot m}\over{m}}{\Bigl\vert}
{\,},
\eqno(6.16)$$
\noindent
provided that 
$2m(t,r)<r<4m(t,r)$.  
In this case, 
the rate of change of the metric with time 
is slow compared to the typical frequencies of the radiation; 
further, 
the time--variation scale of the background space--time metric 
${\gamma}_{{\mu}{\nu}}$ 
is much greater than the period of the waves.  
With frequencies of magnitudes 
${\bigl\vert}k{\bigl\vert}{\;}{\sim}{\;}{m}^{-1}$ 
dominating the radiation, 
and with 
${\bigl\vert}{\dot m}{\bigl\vert}$ 
of order 
$m^{-2}$ 
[61], 
the adiabatic approximation is equivalent to 
$m^{2}\gg{1}{\,}$, 
which corresponds to the semi--classical approximation.  
If, 
as expected [61], 
$m^{3}$ 
is a measure of the time taken by the hole to evaporate, 
then 
$r<4m{\;}{\ll}{\;}{m}^{3}{\,}$, 
provided that 
${m}^{2}\gg{1}{\,}$.  
Thus, 
in the large--$k$ approximation used in deriving Eq.(6.14), 
it is valid at lowest order to neglect time--derivatives 
of the background metric, 
out to radii small compared with the evaporation time 
of the hole and with the time since the hole was formed.\par
\smallskip
\indent
It is natural to define a generalisation 
$r^{*}$ 
of the standard Regge--Wheeler coordinate 
$r^{*}_{~s}$ 
for the Schwarzschild geometry [16,49],
according to 
$${{\partial}\over{\partial{r^{*}}}}
{\;}
=
{\;}
e^{(b-a)/2}{\;}{{\partial}\over{\partial{r}}}
{\;}{\,}.
\eqno(6.17)$$
\noindent
Under the above conditions, 
the time--dependence of 
$r^{*}(t,r)$ 
is negligibly small, 
and one has 
${\,}r^{*}{\;}{\sim}{\;}{r}^{*}_{~s}$ 
for large 
$r{\,}$, 
where, 
by definition,
$$r^{*}_{~s}
{\;} 
=
{\;}
r{\,}
+{\,}2M\log\bigl(\bigl(r/2M\bigr)-1\bigr)
\eqno(6.18)$$
\noindent
is the Regge--Wheeler coordinate, 
expressed in terms of the Schwarzschild radial coordinate 
$r{\,}$.  
In terms of the variable 
${\,}r^{*}{\,}$, 
the approximate 
(adiabatic) 
mode equation (6.14) reads
$${{\partial^{2}{\xi}_{k{\ell}m}}\over{\partial{r^{*2}}}}{\,}
+{\,}\bigl(k^{2}-V_{\ell}\bigr){\,}{\xi}_{k{\ell}m}
{\;} 
=
{\;}
0
{\,}.
\eqno(6.19)$$
\noindent
{\bf 7. Boundary conditions}
\medskip
\indent
We now consider, 
in more detail, 
a set of suitable radial functions
$\{{\xi}_{k{\ell}m}(r)\}$ 
on the final surface 
${\Sigma}_{F}{\;}$.  
As above, 
since the mode equation (6.8) does not depend on the quantum number 
$m{\,}$, 
we may choose 
${\xi}_{k{\ell}m}(r)
=
{\xi}_{k{\ell}}(r)$,
independently of 
$m{\,}$.
\par
\smallskip
\indent
We seek a complete set, 
such that any smooth perturbation field
${\phi}^{(1)}(T,r,{\theta},{\phi})$ 
of rapid decay near spatial infinity,
when restricted to the final surface 
$\{t=T\}$, 
can be expanded in terms of the 
${\xi}_{k{\ell}m}(r)$.  
The 
'left' 
boundary condition on the radial functions 
$\{{\xi}_{k{\ell}}(r)\}$ 
is that of regularity at the origin 
$r
=
0{\;}$:
$${\xi}_{k{\ell}}(0)
{\;}
=
{\;}
0
{\,}.
\eqno(7.1)$$
\noindent
The solution to the radial equation, 
regular near the origin, 
is:
$${\xi}_{k{\ell}}(r)
{\;}{\;}{\,}
=
{\;}{\;}{\,}
r{\,}{\phi}_{k{\ell}}(r)
{\;}{\;}{\,}
\propto
{\;}{\;}{\,}
r{\,}j_{\ell}(kr)
{\;}{\;}{\,}
\propto
{\;}{\;}{\,}
\Bigl\{\bigl({\rm const.}\times(kr)^{{\ell}+1}\bigr) 
+{\,}O\bigl((kr)^{{\ell}+3}\bigr)\Bigr\},
\eqno(7.2)$$ 
\noindent
where, 
again, 
$j_{\ell}$ 
denotes a spherical Bessel function [60]; 
we have assumed that, 
for small 
$r{\,}$, 
one has
$m(r){\;}{\propto}{\;}{r}^{3}{\,}$, 
and we have neglected 
$O(r^{2})$ 
terms.  
These radial functions are purely real, 
for real 
$k$ 
and 
$r{\,}$.  
For 
$k$ 
purely real and positive, 
the radial functions describe standing waves, 
which, 
for mode time--dependence 
$e^{{\pm}ikt}{\,}$,
have equal amounts of 
'ingoing' 
and 
'outgoing' 
radiation.\par
\smallskip
\indent
For the 
'right' 
boundary condition, 
note that the potential
$V_{\ell}(r)$, 
following from Eq.(4.5), 
vanishes sufficiently rapidly as 
$r\longrightarrow{\infty}$ 
that a real solution to Eq.(6.19) obeys
$${\xi}_{k{\ell}}(r)
{\;}{\;}{\;} 
{\sim}
{\;}{\;}{\;}
\Bigl(z_{k{\ell}}{\,}\exp\bigl(ikr^{*}_{~s}\bigr){\,}
+{\,}z^{*}_{k{\ell}}{\,}\exp\bigl(-{\,}ikr^{*}_{~s}\bigr)\Bigr)
\eqno(7.3)$$
\noindent
as 
$r\longrightarrow{\infty}{\,}$.  
Here, 
the 
$z_{k{\ell}}$ 
are certain dimensionless complex coefficients, 
which can be determined 
{\it via} 
the differential equation by using the regularity at 
$r
=
0{\,}$.  
The 
(approximately) 
conserved Wronskian for Eq.(6.19), 
together with Eq.(7.3), 
and the property 
$$\lim_{r^{*}_{~s}\longrightarrow{\infty}} 
{{\exp\bigl(i(k-k^{\prime}){\,}r^{*}_{~s}\bigr)}
\over{\bigl(k-k^{\prime}\bigr)}}
{\;} 
=
{\;}
i{\pi}{\,}{\delta}(k-k'),
\eqno(7.4)$$
\noindent
give the normalisation condition, 
for 
$-{\infty}<k{\,},{\,}k^{\prime}<{\infty}{\,}$
and 
$R_{\infty}\longrightarrow{\infty}{\,}$: 
$$\int^{R_\infty}_{0}dr{\;}{\,}e^{(a-b)/2}{\;}
{\xi}_{k{\ell}}(r){\;}
{\xi}^{*}_{k'{\ell}}(r)\Bigl\arrowvert_{{\Sigma}_{F}}
=
{\;}
2{\pi}{\,}{\bigl\vert}z_{k\ell}{\bigl\vert}^{2}{\,}
\bigl({\delta}(k-k^{\prime})
+{\,}{\delta}(k+k^{\prime})\bigr).
\eqno(7.5)$$
\noindent
This normalisation is only possible within the adiabatic approximation.
Note that the radial functions 
$\{{\xi}_{k{\ell}}\}$ 
form a complete set only for 
$k>0{\,}$, 
as a result of our boundary conditions.\par
\smallskip
\indent
The above result makes it possible to evaluate 
the perturbative massless--scalar contribution 
to the total classical Lorentzian action
$S_{\rm class}{\,}
=
{\,}S^{(0)}_{\rm class}{\,}
+{\,}S^{(2)}_{\rm class}{\,}
+{\,}\ldots{\,}$ 
of Eq.(5.11), 
with 
$S^{(2)}_{\rm class}$ 
given by Eq.(5.12).  
This contribution, 
namely 
$S^{(2)}_{\rm class{\,},{\,}scalar}{\,}$ 
of Eq.(6.3), 
is given in the notation of Eq.(6.7) by 
$$S^{(2)}_{\rm class}\bigl[{\phi}^{(1)}{\,};{\,}T\bigr]
{\;}
=
{\;}
{{1}\over{2}}{\;}
\sum^{\infty}_{{\ell}{\,}={\,}0}{\;}
\sum^{\ell}_{m{\,}={\,}-{\,}{\ell}}{\,}
\int^{R_{\infty}}_{0}dr{\;}{\,}e^{(a-b)/2}{\;}R_{{\ell}m}{\,}
\bigl(\partial_{t}R^{*}_{{\ell}m}\bigr)\bigl\arrowvert_{T}
{\;}{\;},
\eqno(7.6)$$
\noindent
since
$${\int}d{\Omega}{\;}{\,}
Y_{{\ell}m}{\,}Y^{*}_{{\ell}^{\prime}m'}
{\;} 
=
{\;}
{\delta}_{{\ell}{\ell}^{\prime}}{\,}{\delta}_{mm'}
{\;}{\,}.
\eqno(7.7)$$
\noindent
Within the adiabatic approximation above, 
and using Eq.(7.5), 
this gives the frequency--space form of the classical action:
$$S^{(2)}_{\rm class}\bigl[\{a_{k{\ell}m}\}{\,};
{\,}T\bigr]
{\;}
=
{\;}
{\pi}{\,}\sum^{\infty}_{{\ell}{\,}={\,}0}{\;}
\sum^{\ell}_{m{\,}={\,}-{\,}{\ell}}{\,}
\int^{\infty}_{0}dk{\;}{\,}k{\;}
{\bigl\vert}z_{k\ell}{\bigl\vert}^{2}{\;}
{\bigl\vert}a_{k{\ell}m}+{\,}a_{-{\,}k{\ell}m}{\bigl\vert}^{2}{\,}
\cot(kT),
\eqno(7.8)$$
\noindent
in terms of the final data 
$\{a_{k{\ell}m}\}$.
\par
\smallskip
\indent
From a mathematical point of view, 
one would expect to work only with the set 
of square--integrable scalar wave--functions on the final boundary 
${\Sigma}_{F}{\;}$, 
that is, 
the set 
$L^{2}({\Bbb R}^{3}{\,},
{\,}dr{\,}e^{(a-b)/2})$.  
To express this, 
define
$${\psi}_{{\ell}m}(r)
{\;} 
=
{\;}
r{\,}{\int}d{\Omega}{\;}{\,}Y_{{\ell}m}(\Omega){\;}
{\phi}^{(1)}(t,r,{\Omega})\bigl\arrowvert_{t = T}
{\;}{\;}.
\eqno(7.9)$$
\noindent
Then the square--integrability condition reads
$${{1}\over{2{\pi}}}{\,}
\sum_{{\ell}m}{\,}
\int^{R_{\infty}}_{0}dr{\;}{\,}e^{(a-b)/2}{\;}
{\bigl\vert}{\psi}_{{\ell}m}(r){\bigl\vert}^{2}
{\;}{\,}
<
{\;}{\,}
\infty
{\;},
\eqno(7.10)$$
\noindent
or, 
equivalently,
$$\sum_{{\ell}m}{\,}
\int^{\infty}_{-\infty}dk{\;}{\,}
{\bigl\vert}z_{k{\ell}}{\bigl\vert}^{2}{\;}
{\bigl\vert}a_{k{\ell}m}+{\,}a_{-{\,}k{\ell}m}{\bigl\vert}^{2}
{\;}{\,}
<
{\;}{\,}
\infty
{\;}.
\eqno(7.11)$$
\indent
The left--hand sides of Eqs.(7.10,11) are in fact equal. 
This arises from the completeness property
$$e^{(a-b)/2}
\int^{\infty}_{-\infty}dk{\;}{\,}
{{{\xi}_{k{\ell}}(r){\,}{\xi}_{k{\ell}}(r^{\prime})}
\over{{\bigl\vert}z_{k{\ell}}{\bigl\vert}^{2}}}
{\;} 
=
{\;}
4{\pi}{\,}{\delta}(r-{\,}r')
\eqno(7.12)$$
\noindent
and the inverse of Eq.(6.11):
$$a_{k{\ell}m}{\,}+{\,}a_{-{\,}k{\ell}m}
{\;}
=
{\;}
{{1}\over{2{\pi}{\,}{\bigl\vert}z_{k{\ell}}{\bigl\vert}^{2}}}{\,}
\int^{R_{\infty}}_{0}dr{\;}{\,}e^{(a-b)/2}{\;}
{\xi}_{k{\ell}}(r){\,}{\psi}_{{\ell}m}(r).
\eqno(7.13)$$
\smallskip
\indent
From a physical point of view, 
one expects also that taking scalar boundary data 
which are not square--integrable 
will lead to various undesirable properties, 
such as infinite total energy of the system,
or an infinite or ill--defined action.\par
\medskip
\noindent
{\bf 8. Analytic continuation}
\medskip
\indent
The perturbative classical scalar action 
$S^{(2)}_{\rm class}$ 
of Eq.(7.8) was derived subject to the adiabatic approximation 
and to the requirement that the time--interval 
$T$ 
between the initial and final surfaces, 
measured at spatial infinity, 
should be complex, 
of the form 
$T
=
{\mid}T{\mid}\exp(-{\,}i{\theta})$, 
with
${\,}0<{\theta}\leq{\pi}/2{\;}$.  
In this case, 
the term 
$k\cot(kT)$ 
in the integrand of Eq.(7.8) remains bounded near 
$k=0{\,}$,
and one expects to obtain a finite complex--valued action 
$S^{(2)}_{\rm class}[\{a_{k{\ell}m}\}{\,};
{\,}T]$, 
given square--integrable data 
${\phi}^{(1)}$ 
on the final surface 
${\Sigma}_{F}{\;}$.  
Further, 
the dependence of the complex function
$S^{(2)}_{\rm class}[\{a_{k{\ell}m}\}{\,};
{\,}T]$ 
on the complex variable 
$T$ 
is expected to be complex--analytic in this domain 
$(0<{\theta}\leq{\pi}/2)$, 
and, 
following Feynman [14,15], 
ordinary Lorentzian--signature quantum amplitudes 
should be given by the limiting behaviour of 
$\exp(iS^{(2)}_{\rm class})$ 
as 
${\,}{\theta}\longrightarrow{0}_{+}{\;}$.
\par
\smallskip
\indent
If, 
on the other hand, 
one restricts attention to the exactly Lorentzian--signature case 
$({\theta}=0)$, 
then the integral in Eq.(7.8) will typically diverge, 
due to the simple poles on the real--frequency axis at
$$k
{\;}
=
{\;}
k_{n}
{\;}
=
{\;}
{{n{\pi}}\over{T}}
{\qquad}{\qquad}{\qquad}{\quad}
(n=1,2,3,{\,}\ldots{\;}).
\eqno(8.1)$$
\indent
We now restrict attention to the case in which
$T$ 
is only slightly complex, 
writing 
$T
=
{\mid}T{\mid}\exp(-{\,}i{\delta})$,  
with 
${\,}0<{\delta}\ll{1}{\,}$. 
The spherically--symmetric 
'background' 
4--geometry 
${\gamma}_{{\mu}{\nu}}$ 
and scalar field 
${\Phi}$ 
will then be complex. 
Consider an integral such as Eq.(7.8) for 
$S^{(2)}_{\rm class}[\{a_{k{\ell}m}\}{\,};
{\,}T]$.  
Write this as 
$$J
{\;}
=
{\;}
\sum_{{\ell}m}{\,}
\int^{\infty}_{0}dk{\;}{\,}f_{{\ell}m}(k)\cot(kT),
\eqno(8.2)$$ 
\noindent
where 
$$f_{{\ell}m}(k)
{\;} 
=
{\;}
{\pi}k{\,}{\bigl\vert}z_{k{\ell}}{\bigl\vert}^{2}{\;}
{\bigl\vert}a_{k{\ell}m}+{\,}a_{-{\,}k{\ell}m}{\bigl\vert}^{2}
{\,}.
\eqno(8.3)$$   
\noindent
There are infinitely many simple poles of the integrand at 
$k
=
k_{n}{\;}{\,}(n=1,2,{\,}\ldots{\,})$, 
just above the positive real $k$--axis.  
We then deform the original contour 
$C$ 
along the positive real $k$--axis into three parts, 
$C_{\epsilon}{\,},{\,}C_{R}$ 
and 
$C_{\alpha}{\;}$, 
where 
$0<{\alpha}\ll{1}{\,}$.  
The contour 
$C_{\epsilon}$ 
lies in the lower half--plane, 
half--encircling each of the simple poles near the positive real $k$--axis, 
with radius 
${\epsilon}{\,}$.  
The curve 
$C_{R}{\;}$, 
also in the lower half--plane, 
is an arc of a circle 
${\,}{\mid}k{\mid}
=
R$ 
of large radius.  
The curve 
$C_{\alpha}$ 
is part of the radial line 
$\arg(k)
=
{\,}-{\,}{\alpha}{\;}$.  
We write
$$\eqalign{J
{\;}&
=
{\;}
\sum_{{\ell}m}{\,}
\int_{{C_{\alpha}}+C_{R}-C_{\epsilon}}
dk{\;}{\,}f_{{\ell}m}(k){\,}\cot(kT)\cr
&=
{\;}
J_{\alpha}+{\,}J_{R}+{\,}J_{\epsilon}
{\;}{\,}.\cr}
\eqno(8.4)$$
\indent
Starting with the integral 
$J_{R}{\;}$, 
one finds 
$${\bigl\vert}J_{R}{\bigl\vert}
{\;}{\,} 
\leq
{\;}{\,}
\sum_{{\ell}m}{\,}
\int^{\alpha}_{0}d{\theta}{\;}{\,}R{\;}
{\bigl\vert}f_{{\ell}m}(R,{\theta}){\bigl\vert}{\;}
\coth\bigl({\bigl\vert}T{\bigl\vert}R{\,}\sin{\theta}\bigr),
\eqno(8.5)$$
\noindent
where 
$k
=
Re^{-{\,}i{\theta}}$ 
on 
$C_{R}{\;}$, 
and we have used 
${\mid}\cot(kT){\mid}
\leq
\coth\bigl({\mid}T{\mid}R\sin{\theta}\bigr)$.  
One expects that,
when the limit
$R\longrightarrow{\infty}$ 
is eventually taken, 
the contribution from 
$C_{R}$ 
to the total action should vanish; 
this requires that 
${\mid}f_{{\ell}m}(k){\mid}$ 
should decay at least as rapidly as 
${\,}{\mid}k{\mid}^{-2}$, 
as
${\,}{\mid}k{\mid}\longrightarrow{\infty}{\,}$.  
In fact, 
on dimensional grounds, 
one expects that
$${\bigl\vert}f_{{\ell}m}(k){\bigl\vert}
{\;}{\;}{\;}{\,}
{\sim}
{\;}{\;}{\;}{\,}
{\rm const.}{\;}{\times}{\;}{\bigl\vert}k{\bigl\vert}^{-3}
\eqno(8.6)$$
\noindent
as 
${\,}{\mid}k{\mid}\longrightarrow{\infty}{\,}$.  
To see this, 
rewrite the radial equation (6.19) in terms of the operator
$${\cal{L}}_{\ell}
{\;} 
=
{\;}
e^{(b-a)/2}{\;}
{{d}\over{dr}}
\Bigl(e^{(b-a)/2}{{d}\over{dr}}\bigl({\;}{\,}\bigr)\Bigr) 
-{\,}V_{\ell}(r),
\eqno(8.7)$$
which is self--adjoint with respect to the inner product in Eq.(7.5).  
Then note that Eq.(7.13) can be rewritten as
$$a_{k{\ell}m}{\,}+{\,}a_{-{\,}k{\ell}m}
{\;} 
=
{\;}
{{-{\,}1}\over{2{\pi}k^{2}{\,}
{\bigl\vert}z_{k{\ell}}{\bigl\vert}^{2}}}{\,}
\int^{R_{\infty}}_{0}dr{\;}{\,}
e^{(a-b)/2}{\;}{\xi}_{k{\ell}}(r){\;}
{\cal L}_{\ell}{\psi}_{{\ell}m}(r).
\eqno(8.8)$$
\noindent
We have used the boundary condition (7.1) and assumed that 
${\psi}_{{\ell}m}(r)$ 
dies out at large 
$r{\,}$.  
The form (8.8) is just an expression of the self--adjointness 
of the radial equation.  
Now consider the dimensions of the quantities involved.  
One finds [1] that 
${\psi}_{{\ell}m}(r)$ 
has dimensions of length and that
${\mid}z_{k\ell}{\mid}^{2}$ 
is dimensionless.  
In the limit 
$R_{\infty}\longrightarrow{\infty}{\,}$, 
and for large 
${\,}k{\,}$ 
(so taking a WKB approximation for the radial functions), 
the integral in Eq.(8.8) can only involve the dimensionless frequency 
$2Mk{\;}$, 
where 
$M$ 
is the total mass 
(true 
$ADM$ 
mass) 
of the space--time. 
This gives the desired behaviour (8.6) at large 
${\mid}k{\mid}{\,}$.
\par
\smallskip
\indent
The contour 
$C_{\epsilon}$ 
gives a purely imaginary contribution to the total Lorentzian action; 
also 
(see below), 
the curve 
$C_{\alpha}$ 
gives a complex contribution.  
We shall interpret the quantity  
$\exp[-{\,}2{{\,}\rm{Im}}(S)]$, 
up to normalisation, 
as describing the conditional probability density 
over the final boundary data.  
To compute 
$J_{\epsilon}{\;}$, 
we assume that 
$f_{{\ell}m}(k)$ 
is analytic in a neighbourhood of 
${\,}k
=
{\sigma}_{n}{\;}$, 
where
$${\sigma}_{n}
{\;}
=
{\;}
{{n{\pi}}\over{{\mid}T{\mid}}}
{\qquad}{\qquad}{\qquad}{\quad}
(n=1,2,3,{\,}\ldots{\,}).
\eqno(8.9)$$
\noindent
Note the difference between the definitions (8.1) of
$k_{n}$
and (8.9) of
${\sigma}_{n}{\;}$.  
Then
$$\eqalign{J_{\epsilon}
{\;}
&=
{\,}
-{\,}\lim_{{\epsilon}{\,}\longrightarrow{\,}0}{\,}
\sum_{{\ell}m}{\,}
\int_{C_{\epsilon}}
dk{\;}{\,}f_{{\ell}m}(k){\,}\cot\bigl(k{\mid}T{\mid}\bigr)\cr
&=
{\;}
{{i{\pi}}\over{{\mid}T{\mid}}}{\;}
\sum_{{\ell}m}{\,}
\sum^{\infty}_{n{\,}={\,}1}{\,}
f_{{\ell}m}({\sigma}_{n}).\cr}
\eqno(8.10)$$
\indent
For the curve 
$C_{\alpha}{\,}$, 
one has 
$$J_{\alpha}
{\;} 
=
{\,}
-{\,}\sum_{{\ell}m}{\,}
\int^{R}_{0}d{\mid}k{\mid}{\;}{\,}
e^{-{\,}i{\alpha}}{\;}
f_{{\ell}m}\bigl({\mid}k{\mid},{\alpha}\bigr){\,}
\cot\bigl({\mid}k{\mid}{\,}e^{-{\,}i{\alpha}}{\,}{\mid}T{\mid}\bigr).
\eqno(8.11)$$ 
\noindent
We shall need the properties [60]
$$\cot(x)
{\;} 
=
{\,}
\sum^{\infty}_{n{\,}={\,}-{\,}\infty}{\,}
{{1}\over{(x-{\,}n{\pi})}}
\eqno(8.12)$$
\noindent
and
$${{1}\over{(x-{\,}a{\,}\pm{\,}i{\epsilon})}}
{\;}
=
{\;}{\,}
{\rm P.P.}{\;}{{1}\over(x-{\,}a)}{\;}
\mp{\;}i{\pi}{\,}{\delta}(x-{\,}a),
\eqno(8.13)$$
\noindent
where 
${\rm P.P.}$ 
denotes the principal part.  
Assuming that 
$f_{{\ell}m}(k)$ 
is regular along 
$C_{\alpha}{\;}$, 
one has, 
for small 
${\alpha}{\;}$:
$$J^{(1)}_{\alpha}
{\;} 
=
{\;}
\lim_{{\alpha}{\,}\rightarrow{\,}0_{+}}
(1-{\,}i{\alpha}){\,}
\sum_{{\ell}m}{\,}
\sum^{\infty}_{n{\,}={\,}-{\,}\infty}{\,}
\int^{R}_{0}d{\mid}k{\mid}{\;}
{{f_{{\ell}m}\bigl({\mid}k{\mid}{\,},{\alpha}\bigr)}
\over{\bigl({\mid}kT{\mid}-n{\pi}-i{\alpha}\bigr)}}
\eqno(8.14)$$
\noindent
In the further limit 
$R\longrightarrow{\infty}{\,}$, 
this gives
$$J_{\alpha}
{\;} 
=
{\;}{\,}
{\rm P.V.}{\,}
+{\;}{{i{\pi}}\over{{\mid}T{\mid}}}{\;}
\sum_{{\ell}m}{\,}
\sum^{\infty}_{n{\,}={\,}1}{\,}
f_{{\ell}m}({\sigma}_{n}),
\eqno(8.15)$$
\noindent
where 
${\rm P.V.}$ 
denotes the principal--value part of the integral.\par
\smallskip
\indent
Using Eqs.(8.4,5,10,15), 
the classical action for massless scalar--field perturbations, 
when
$T
=
{\mid}T{\mid}\exp(-{\,}i{\delta})$ 
is very slightly complex, 
is
$$\eqalign{S^{(2)}_{\rm class}
\bigl[\{a_{k{\ell}m}\}{\,};
{\,}{\mid}T{\mid}\bigr]
{\;}
&=
{\;}
{\rm real{\;}{\,}part}{\,} 
+{\;}{{2i{\pi}}\over{{\mid}T{\mid}}}{\;}
\sum^{\infty}_{{\ell}{\,}={\,}0}{\;}
\sum^{\ell}_{m{\,}={\,}-{\,}{\ell}}{\;}
\sum^{\infty}_{n{\,}={\,}1}{\,}
f_{{\ell}m}({\sigma}_{n})\cr
&=
{\;}
{\rm real{\;}{\,}part}{\,}
+{\;}{{2i{\pi}^{2}}\over{{\mid}T{\mid}}}{\;}
\sum_{{\ell}mn}{\,}
{\sigma}_{n}{\;}
{\bigl\vert}z_{n{\ell}}{\bigl\vert}^{2}{\;}
{\bigl\vert}a_{n{\ell}m}+a_{-{\,}n{\ell}m}{\bigl\vert}^{2}
{\,}.\cr}
\eqno(8.16)$$
\noindent
The real part of 
$S^{(2)}_{\rm class}{\,}$ 
is, 
of course, 
also calculable from the equations above.  
It contains the principal--value term and the real part of Eq.(8.10). 
The main, 
semi--classical, 
contribution to the quantum amplitude is then 
$\exp(iS^{(2)}_{\rm class}[\{a_{k{\ell}m}\}{\,};
{\,}{\mid}T{\mid}])$.  
The probability distribution for final configurations involves only 
${\rm Im}(S^{(2)}_{\rm class})$;
the more probable configurations have 
$S^{(2)}_{\rm class}$ 
lying only infinitesimally in the upper half--plane.  
Whether probable or not, 
those final configurations  
$\{a_{k{\ell}m}\}$ 
which contribute to the probability distribution 
must yield finite expressions in the infinite sums over 
$n{\,}{\ell}$ 
in Eq.(8.16).  
There will be a corresponding restriction when the data 
are instead described in terms of the spatial configurations 
$\{{\psi}_{{\ell}m}(r)\}$. 
Also, 
as can be seen in [5], 
the complex quantities 
$z_{n{\ell}}{\,}(a_{n{\ell}m}+{\,}a_{-{\,}n{\ell}m})$ 
appearing in Eq.(8.16) are related to Bogoliubov transformations 
between initial and final states, 
thus providing a further characterisation of the finiteness of 
${\rm Im}(S^{(2)}_{\rm class})$ 
in Eq.(8.16).\par
\smallskip
\indent
With regard to the sum over 
${\ell}$ 
in Eq.(8.16), 
one imagines that a cut--off 
${\ell}_{{\rm max}}$ 
can be provided by the radial equation (6.19).
In the region where 
$(V_{\ell}(r)-k^{2})>0{\,}$, 
one has exponentially growing radial functions, 
whereas for 
$(V_{\ell}(r)-k^{2})<0$ 
one has oscillatory radial functions. 
One defines 
${\ell}_{\rm max}$ 
by
$(V_{{\ell}{\rm max}}(r)-k^{2})
=
0$ 
and restricts attention mainly to oscillatory solutions.\par
\smallskip
\indent
When one has both initial and final non--zero Dirichlet data labelled by 
'coordinates' 
$\{a^{(I)}_{k{\ell}m}\}$ 
and 
$\{a^{(F)}_{k{\ell}m}\}$, 
the perturbative classical scalar action 
$S^{(2)}_{\rm class}$ 
includes separate terms of the form (8.16) for the initial and final data.  
But
$S^{(2)}_{\rm class}$ 
also includes a cross--term between
$a^{(I)}_{k{\ell}m}$ 
and 
$a^{(F)}_{k{\ell}m}{\;}$, 
which represents the correlation or mixing 
between the initial and final data.  
The total action will naturally be symmetric in 
$a^{(I)}_{k{\ell}m}$ 
and
$a^{(F)}_{k{\ell}m}{\;}$, 
and the coefficients 
$z_{n{\ell}}$ 
will be the same 
(they are time--independent) 
up to a phase.  
For large 
${\mid}T{\mid}{\,}$,
the cross--term becomes negligible, 
and one has two independent contributions to the classical action, 
one being a functional of
$\{a^{(I)}_{k{\ell}m}\}$, 
the other of 
$\{a^{(F)}_{k{\ell}m}\}$.
\par
\medskip
\noindent
{\bf 9. Conclusion}
\medskip
\indent 
In this paper, 
we have derived through Eq.(8.16) the quantum amplitude 
for a spheric-ally--symmetric configuration 
$(h_{ij}{\,},
{\phi})_{I}$
on the initial surface 
${\Sigma}_{I}$ 
to become a configuration 
$(h_{ij}{\,},
{\phi})_{F}$ 
on the final surface 
${\Sigma}_{F}{\;}$, 
with Lorentzian time--interval 
$T$ 
at spatial infinity. 
Here, 
${\phi}_{F}$ 
will,
in general, 
be anisotropic, 
although 
(for simplicity) 
we assumed that the final 3--dimensional metric 
$h_{ijF}$ 
is also spherically symmetric.  
In the amplitude, 
which for a locally--supersymmetric theory
is proportional to 
$\exp(iS^{(2)}_{\rm class})$, 
the classical action depends approximately quadratically on the 
(non--spherical) 
perturbative part of the final data 
${\phi}_{F}{\;}$.
Further, 
$S^{(2)}_{\rm class}$ 
has both a real and an imaginary part.  
The imaginary part leads to a Gaussian probability density
${\mid}{\Phi}{\mid}^{2}
\propto
\exp(-{\,}2{\,}{\rm Im}(S^{(2)}_{\rm class}))$, 
while the real part gives rapid oscillations in the phase 
of the quantum amplitude or wave function 
${\Phi}{\,}$.
\par
\smallskip
\indent
We have arrived at a quantum amplitude 
(not just a probability distribution) 
for such processes, 
simply by following Feynman's 
$+{\,}i{\epsilon}$ 
prescription,  
applied to the exactly semi--classical expression (2.2)
for the quantum amplitude.
This,
in turn,
is derived
{\it via}
Dirac's canonical--quantisation approach,
for a locally--supersymmetric Lagrangian such as that of gauge--invariant 
$N=1$
supergravity.
The boundary conditions are treated by rotating the time--interval 
$T$ 
into the lower--half complex plane:
$T\longrightarrow{\mid}T{\mid}\exp(-{\,}i{\theta})$, 
for 
$0<{\theta}\leq{\pi}/2{\;}$.  
We then studied the classical 
and corresponding quantum--mechanical boundary--value problems, 
before rotating 
${\,}{\theta}{\,}$ 
back towards zero.\par
\smallskip
\indent
These ideas have also been applied to black--hole evaporation 
for particles of spin 
$1$ 
and 
$2{\,}$ 
[2], 
and to the fermionic spin--${{1}\over{2}}$ case [3].
But,
in those references,
the form of the complex quantum amplitudes 
was not derived in the greater detail given in the full computation 
of spin--0 amplitudes of the present paper. 
In [5,6], 
we made a connection relating the present description 
and calculation of quantum amplitudes to the familiar description 
in terms of Bogoliubov coefficients [62--64].  
A more general conceptual framework has been provided within the language 
of coherent and squeezed states [65,66].\par
\medskip
\parindent = 1 pt
\noindent
{\bf References}
\medskip
\indent [1]  A.N.St.J.Farley, 
'Quantum Amplitudes in Black--Hole Evaporation', 
Cambridge Ph.D. dissertation, approved 2002 (unpublished).\par

\indent [2]  A.N.St.J.Farley and P.D.D'Eath, 
{~}'Quantum amplitudes in black--hole evaporation: 
Spins 1 and 2',
{~}Ann. Phys. (N.Y.) {\bf 321} 1334 {~}(2006).
{~}(arXiv gr--qc/0708.2013)\par

\indent [3] A.N.St.J.Farley {~}and {~}P.D.D'Eath, 
{~~~~~}'Spin--${{1}\over{2}}$ amplitudes 
in black--hole evaporation', 
{~}Class. Quantum Grav. {\bf 22} 3001 {~}(2005).
{~}(arXiv gr--qc/0510036)\par

\indent [4] A.N.St.J.Farley and P.D.D'Eath, 
{~}'Scalar--field amplitudes in black--hole evaporation',
{~}Phys. Lett. B {\bf 601} 184 {~}(2004).
{~}(arXiv gr--qc/0407086)\par    

\indent [5] A.N.St.J.Farley and P.D.D'Eath, 
{~}'Bogoliubov transformations for amplitudes 
in black--hole evaporation',
{~}Phys. Lett. B {\bf 613} 181 {~}(2005). 
{~}(arXiv gr--qc/0510027)\par

\indent [6] A.N.St.J.Farley and P.D.D'Eath. 
{~}'Bogoliubov transformations 
in black--hole evaporation',
{~}Int. J. Mod. Phys. D {\bf 16} 569 {~}(2007).
{~}(arXiv gr--qc/0510043).\par

\indent [7] M.K.Parikh and F.Wilczek, 
{~}Phys. Rev. Lett. {\bf 85} 5042 {~}(2000).\par

\indent [8] M.Parikh, 
{~}Gen. Relativ. Gravit. {\bf 36} 2419 {~}(2004).\par 

\indent [9]  P.R.Garabedian, 
{~}{\it Partial Differential Equations} 
{~}(Wiley, New York) {~}(1964).\par

\indent [10] Y.Choquet--Bruhat {~}and {~}J.W.York, 
{~~}'The Cauchy Problem', 
{~}in 
{~}{\it  General {~}Relativity {~}and {~}Gravitation}, 
{~}ed. A.Held, {~}(Plenum, New York) 
{~}Vol.1, {~}p.99 {~}(1980).\par

\indent [11] A.E.Fischer {~}and {~}J.E.Marsden, 
{~~~~}'The initial value problem 
and the dynamical formulation of general relativity', 
{~}in 
{~~}{\it General {~}Relativity}, 
{~~}eds. S.W.Hawking {~}and {~}W.Israel, 
{~~}(Cambridge University Press, {~}Cambridge) 
{~}p.138 {~}(1979).\par

\indent [12] P.D.D'Eath, 
{~~}{\it Supersymmetric {~}Quantum {~}Cosmology} 
{~~}(Cambridge University Press, Cambridge) {~}(1996).\par

\indent [13] W.McLean, 
{~~~}{\it Strongly {~}Elliptic {~}Systems 
{~}and {~}Boundary {~}Integral {~}Equations}, 
{~~~}(Cambridge University Press, {~}Cambridge) {~}(2000).\par 

\indent [14] R.P.Feynman {~}and {~}A.R.Hibbs, 
{~~~~~}{\it Quantum {~}Mechanics 
{~}and {~}Path {~}Integrals} 
{~~~~}(McGraw--Hill, {~}New York) {~}(1965).\par

\indent [15] C.Itzykson {~}and {~}J.--B.Zuber,
{~}{\it Quantum Field Theory} {~}(McGraw--Hill, 
{~}New York) {~}(1980).\par

\indent [16] C.W.Misner, K.S.Thorne and J.A.Wheeler, 
{\it Gravitation} (Freeman, San Francisco) (1973).\par

\indent [17] J.A.Wheeler, 
{~~~}'Superspace {~}and {~}the {~}Nature 
{~}of {~}Quantum {~}Geometrodynamics' 
{~~~}in 
{~~}{\it Battelle {~}Rencontres}, 
{~}eds. C.M.DeWitt {~}and {~}J.A.Wheeler 
{~}(W.A.Benjamin, {~}New York) {~}p.303 {~}(1968).\par

\indent [18] P.D.D'Eath, 
{~}Phys. Rev. D {\bf 24} 811 {~}(1981).\par

\indent [19] P.C.Vaidya,  
{~}Proc. Indian Acad. Sci. {\bf A 33} 264 {~}(1951).\par

\indent [20] R.W.Lindquist, {~}R.A.Schwartz {~}and {~}C.W.Misner, 
{~~}Phys. Rev. {\bf 137} 1364 {~}(1965).\par

\indent [21] A.N.St.J.Farley {~}and {~}P.D.D'Eath, 
{~~~}'Vaidya space--time in black--hole evaporation',
{~~}Gen. Relativ. Gravit. {\bf 38} 425 {~}(2006).
{~}(arXiv gr--qc/0510040)\par
 
\indent [22] P.D.D'Eath, 
{~~}'Loop {~}amplitudes {~}in {~}supergravity 
{~}by {~}canonical {~}quantization', 
{~~}in 
{~}{\it Fundamental {~}Problems 
{~}in {~}Classical, {~}Quantum {~}and {~}String {~}Gravity}, 
{~~}ed. N.S\'anchez {~}(Observatoire de Paris) 
{~}p.166 {~}(1999).
{~}(arXiv hep--th/9807028).\par 

\indent [23] P.D.D'Eath, 
{~}'What {~}local {~}supersymmetry 
{~}can {~}do {~}for {~}quantum {~}cosmology', 
{~~}in 
{~~}{\it The {~}Future 
{~}of {~}Theoretical {~}Physics {~}and {~}Cosmology}, 
{~~}eds. G.W.Gibbons, {~}E.P.S.Shellard {~}and {~}S.J.Rankin 
{~}(Cambridge University Press, {~}Cambridge) 
{~}p.693 {~}(2003)
{~}(arXiv gr--qc/0511042).\par

\indent [24] P.D.D'Eath, 
'Dirac quantization of $N=1$ supergravity 
leads to semi--classical amplitudes', 
{~~}(unpublished).\par

\indent [25] J.B.Hartle {~}and {~}S.W. Hawking, 
{~}Phys. Rev. D {\bf 28} 2960 {~}(1983).\par

\indent [26]  P.A.M.Dirac, 
{~~}{\it Lectures {~}on {~}Quantum {~}Mechanics} 
{~}(Academic Press, {~}New York) {~}(1965).\par

\indent [27] J.Wess {~}and {~}J.Bagger, 
{~~~~~~~}{\it Supersymmetry {~}and {~}Supergravity} 
{~~~~}2nd. edition, 
{~~~~}(Princeton University Press, {~}Princeton) 
{~}(1992).\par

\indent [28] P.D.D'Eath, 
{~~}{\it Black {~}Holes: {~}Gravitational {~}Interactions} 
{~}(Oxford University Press, {~}Oxford) {~}(1996).\par

\indent [29]  G. 't Hooft, 
{~}Phys. Lett. B {\bf 198} 61 {~}(1987).\par

\indent [30]  S. Giddings, 
{~}'Black {~}holes {~}at {~}accelerators',
{~~}in 
{~}{\it The {~}Future 
{~}of {~}Theoretical {~}Physics {~}and {~}Cosmology}, 
{~~}eds. G.W.Gibbons, {~}E.P.S.Shellard {~}and {~}S.J.Rankin 
{~}(Cambridge University Press, {~}Cambridge) 
{~}p.278 {~}(2003).\par

\indent [31] A.N.St.J.Farley {~}and {~}P.D.D'Eath, 
{~}'Spin--${{3}\over{2}}$ amplitudes 
in black--hole evaporation', 
{~~}in progress.\par 

\indent [32] J.Mathews, 
{~}J. Soc. Ind. Appl. Math. {\bf 10} 768 {~}(1962).\par 

\indent [33] J.N.Goldberg, {~}A.J.MacFarlane, 
{~}E.T.Newman, {~}F.Rohrlich {~}and {~}E.C.G.Sudarshan, 
{~}J. Math. Phys. {\bf 8} 2155 {~}(1967).\par

\indent [34] S.Kobayashi {~}and {~}K.Nomizu, 
{~~~}{\it Foundations {~}of {~}Differential {~}Geometry}, 
{~~}Vol. II 
{~~}(Wiley, {~}New York) {~}(1969).\par

\indent [35] P.D.D'Eath {~}and {~}A.Sornborger, 
{~}Class. Quantum Grav. {\bf 15} 3435 {~}(1998).\par

\indent [36] D.Christodoulou, 
{~~}Commun. {~}Math. {~}Phys. {\bf 105} {~}337 {~}(1986); 
{~}{\bf 106} {~}587 {~}(1986); 
{~~}{\bf 109} 591, 613 {~}(1987);
{~~}Commun. Pure Appl. Math. {~}{\bf 44} {~}339 {~}(1991); 
{~}{\bf 46} {~}1131 {~}(1993).\par 

\indent [37] R.Geroch, 
{~}Commun. Math. Phys. {\bf 13} 180 {~}(1969).\par 

\indent [38] P.D.D'Eath, 
{~~}'Numerical and analytic estimates 
for Einstein/scalar boundary--value problems', 
{~}in progress.\par

\indent [39] M.W.Choptuik,
{~~~~}' 'Critical' Behaviour 
in Massless Scalar Field Collapse', 
{~~~~}in 
{~}{\it Approaches {~}to {~}Numerical {~}Relativity}, 
{~~}ed. R.d'Inverno 
{~~}(Cambridge University Press, {~}Cambridge) 
{~}(1992).\par

\indent [40] M.W.Choptuik, 
{~}Phys. Rev. Lett. {\bf 70} 9 {~}(1993).\par 

\indent [41] S.W.Hawking {~}and {~}G.F.R.Ellis, 
{~~~~}{\it The {~}large {~}scale {~}structure 
{~}of {~}space--time} 
{~~}(Cambridge University Press, {~}Cambridge) {~}(1973).\par

\indent [42] U.H.Gerlach {~}and {~}U.K.Sengupta, 
{~}Phys. Rev. D {\bf 18} 1789 {~}(1978).\par

\indent [43] D.Brill {~}and {~}J.B.Hartle,  
{~}Phys. Rev. {\bf 135} 1327 {~}(1964).\par

\indent [44] R.Isaacson, 
{~}Phys. Rev. {\bf 166} 1263, 1272 {~}(1968).\par

\indent [45] A.Nayfeh, {~}{\it Perturbation {~}Methods} 
{~}(Wiley--Interscience, {~}New York) {~}(1973).\par

\indent [46] C.M.Bender {~}and {~}S.A.Orszag, 
{~~}{\it Advanced {~}Mathematical {~}Methods 
{~}for {~}Scientists {~}and {~}Engineers} 
{~~}(Springer, {~}New York) {~}(1999).\par

\indent [47] R.d'Inverno, 
{~~}{\it Introducing {~}Einstein's {~}Relativity}
{~}(Oxford University Press, {~}Oxford) {~}(1992).\par

\indent [48] J.D.Jackson, 
{~}{\it Classical {~}Electrodynamics} 
{~}(Wiley, {~}New York) {~}(1975).\par

\indent [49] T.Regge {~}and {~}J.A.Wheeler, 
{~}Phys. Rev. {\bf 108} 1063 {~}(1957).\par

\indent [50] C.V.Vishveshwara, 
{~}Phys. Rev. D {\bf 1} 2870 {~}(1970).\par

\indent [51] F.J.Zerilli, 
{~}Phys. Rev. D {\bf 2} 2141 {~}(1970).\par

\indent [52] J.A.H.Futterman, {~}F.A.Handler 
{~}and {~}R.A.Matzner,   
{~~}{\it Scattering {~}from {~}Black {~}Holes}
{~~}(Cambridge University Press, {~}Cambridge) {~}(1988).\par

\indent [53] R.Arnowitt, {~}S.Deser {~}and {~}C.W.Misner, 
{~~}'Dynamics {~}of {~}General {~}Relativity', 
{~~}in 
{~~}{\it Gravitation: An Introduction to Current Research}, 
{~}ed. L.Witten {~}(Wiley, New York) {~}(1962).\par

\indent [54]  P.H\'aj\'i\v cek {~}and {~}W.Israel, 
{~}Phys. Lett. A {\bf 80} 9 {~}(1980).\par

\indent [55] J.Bardeen, 
{~}Phys. Rev. Lett. {\bf 46} 382 {~}(1981).\par

\indent [56]  W.A.Hiscock, 
{~}Phys. Rev D {\bf 23} 2813, 2823 {~}(1981).\par

\indent [57] G.W.Gibbons {~}and {~}S.W.Hawking, 
{~}Phys. Rev. D {\bf 15} 2738 {~}(1977).\par

\indent [58] J.J.Halliwell {~}and {~}S.W.Hawking, 
{~}Phys. Rev. D {\bf 31} 1777 {~}(1985).\par

\indent [59]  R.M.Wald, 
{~}Phys. Rev. D {\bf 13} 3176 {~}(1976).\par

\indent [60] M.Abramowitz {~}and {~}I.A.Stegun, 
{~~}{\it Handbook {~}of {~}Mathematical {~}Functions} 
{~~}(Dover, New York) {~}(1964).\par

\indent [61] D.N.Page {~}and {~}S.W.Hawking, 
{~}Astrophys. J. {\bf 206} 1 {~}(1976).\par

\indent [62] S.W.Hawking,  
{~}Commun. Math. Phys. {\bf 43} 199 {~}(1975).\par

\indent [63] N.D.Birrell {~}and {~}P.C.W.Davies, 
{~~}{\it Quantum {~}fields {~}in {~}curved {~}space} 
{~~}(Cambridge University Press, {~}Cambridge) {~}(1982).\par

\indent [64] V.P.Frolov {~}and {~}I.D.Novikov,
{~}{\it Black Hole Physics} 
{~}(Kluwer Academic, {~}Dordrecht) {~}(1998).\par

\indent [65] A.N.St.J.Farley {~}and {~}P.D.D'Eath, 
{~~}'Coherent and squeezed states 
in black--hole evaporation',
{~~}Phys. Lett. B {\bf 634} 419 {~}(2006).
{~}(arXiv gr--qc/0603092)\par

\indent [66] A.N.St.J.Farley and P.D.D'Eath, 
{~}'Quantum amplitudes in black--hole evaporation: 
coherent and squeezed states', 
{~~}Class. Quantum Grav. {\bf 24} 105 {~}(2006).
{~}(arXiv gr--qc/0708.2018)\par

\end